\documentclass[a4paper,twocolumn,11pt,accepted=2024-07-14]{quantumarticle}
\pdfoutput=1
\usepackage[utf8]{inputenc}
\usepackage[english]{babel}
\usepackage[T1]{fontenc}
\usepackage{amsmath}
\usepackage{hyperref}
\usepackage{soul}
\usepackage{subcaption}
\usepackage{xcolor}
\usepackage{soul}

\usepackage{lipsum}
\usepackage{tikz}
\usetikzlibrary{quantikz2}
\usepackage{csquotes}
\usepackage{adjustbox}

\begin{document}

\title{Entanglement spectrum of matchgate circuits with universal and non-universal resources}

\author{Andrew M. Projansky}
\affiliation{Department of Physics and Astronomy, Dartmouth College, Hanover, New Hampshire, USA 03755}
\email{andrew.m.projansky.gr@dartmouth.edu}
\author{Joshuah T. Heath}
\affiliation{Department of Physics and Astronomy, Dartmouth College, Hanover, New Hampshire, USA 03755}
\author{James D. Whitfield}
\affiliation{Department of Physics and Astronomy, Dartmouth College, Hanover, New Hampshire, USA 03755}
\affiliation{AWS Center for Quantum Computing, Pasadena, California, USA 91125}
\email{james.d.whitfield@dartmouth.edu}

\maketitle

\begin{abstract}
    \noindent {The entanglement level statistics of a quantum state have recently been proposed to be a signature of universality in the underlying quantum circuit. This is a consequence of level repulsion in the entanglement spectra being tied to the integrability of entanglement generated. However, such studies of the level-spacing statistics in the entanglement spectrum have thus far been limited to the output states of Clifford and Haar random circuits on product state inputs. In this work, we provide the first example of a circuit which is composed of a simulable gate set, yet has a Wigner-Dyson distributed entanglement level spectrum without any perturbing universal element. We first show that, for matchgate circuits acting on random product states, Wigner-Dyson statistics emerge by virtue of a single SWAP gate, in direct analog to previous studies on Clifford circuits. We then examine the entanglement spectrum of matchgate circuits with varied input states, and find a sharp jump in the complexity of entanglement as we go from two- to three-qubit entangled inputs. Studying Clifford and matchgate hybrid circuits, we find examples of classically simulable circuits whose output states exhibit Wigner-Dyson entanglement level statistics in the absence of universal quantum gate elements. Our study thus provides strong evidence that entanglement spectrum is not strongly connected to notions of simulability in any given quantum circuit.}
\end{abstract}

\section{Introduction}
\label{sec:intro}
The question of integrability (or a lack thereof) in a closed quantum system may be answered by studying the statistical distribution of energy eigenvalues \cite{DOIKOU_2010, Scaramazza_2016, Gubin_2012, Rabson_2004}.
Most integrable systems are characterized by Poisson-distributed energy level spacing, while non-integrable systems instead exhibit energy level repulsion and are hence characterized by the emergence of a Wigner-Dyson distributed energy spectrum.
{\color{black}This follows from the Bohigas-Giannoni-Schmit (BGS) conjecture \cite{PhysRevLett.52.1}, which postulates that the statistical spectrum of Hamiltonian whose classical limit is chaotic may be modeled with random matrices chosen from a Gaussian ensemble such as the GOE or GUE ensembles.} {\color{black}Nevertheless, it is important to note that having a model described by matrices from one of these ensembles does not necessarily mean that model is chaotic. There exist examples of non-chaotic systems with Wigner-Dyson energy level statistics \cite{PhysRevA.42.2431}.} \\
\indent In recent years,
the relationship between chaos and spectra has been carried to the field of quantum information and entanglement dynamics{\color{black}. In addition to the level statistics of energy eigenvalues,} we may {\color{black}also} consider {\it entanglement} level statistics of some time-evolved state, where the entanglement spectrum is defined as the distribution of  eigenvalues of a reduced density matrix \cite{Li_2008, Zhang_2020, stabspec, True_2022, tirrito2023quantifying}. 

Integrability of entanglement is defined in Shaffer et al.\ \cite{entspecirrev} in the context of the reversibility of entanglement under a Metropolis-like Monte Carlo algorithm \cite{irrev}. 
The authors find that states which exhibit Poisson-distributed level statistics exhibit reversible entanglement, while states that exhibit Wigner-Dyson distributed entanglement spectra cannot be reversed \cite{entspecirrev}. As a consequence, the authors of \cite{irrev} suggest that signatures of irreversibility of entanglement, characterized by the level statistics of entanglement spectra, indicate a \textit{complexity of entanglement} in states of interest. Said otherwise, in their opinion, \enquote{... irreversibility is a complexity problem} \cite{entspecirrev}. \\
\indent In this paper, we will discuss the \textit{complexity (or integrability) of entanglement} in a similar fashion; i.e., through the lens of the entanglement spectral statistics. Integrability of entanglement is not only connected to the irreversibility of entanglement, but has also been shown to be an indicator for some universal quantum gate element in the circuit. Within the context of random Clifford circuits, a single $T$ gate is sufficient to change the Poisson-distributed entanglement spectrum of a Clifford circuit on product state inputs to be Wigner-Dyson distributed \cite{ShiyuChamon}. Entanglement spectrum statistics are therefore often seen as a promising indicator for some universal quantum gate element or some non-simulable quantum circuit task. \\
\indent In the present literature, rigorous comparisons have been made between the entanglement spectrum generated by Clifford circuits and Haar random circuits \cite{stabspec, irrev, entspecirrev} to study randomization and $t$-designs with Cliffords and $T$ gates. In this paper, we focus not on the connection between entanglement spectrum statistics and randomization, but on the connection between entanglement spectrum statistics and simulability. Instead of remaining in the context of Clifford circuits, we instead turn our attention to matchgate and hybrid matchgate with Clifford circuits. Matchgate circuits define a continuous family of gates related to free fermion evolution which are simulable on arbitrary product state inputs \cite{TerhalDivincenzo, JozsaMiyake, Brod_2016, MatchwOutside}. In this way, matchgates supply us with a unique opportunity to study entanglement dynamics and simulability in a family of quantum circuits distinct from the Cliffords \cite{JoszaCliffC, MatchwOutside}. { \color{black}Matchgates also allow us to connect entanglement spectra statistics to more physically-motivated fermionic models. This is in contrast to the entanglement spectra of Clifford plus $T$ gates, which do not enjoy as clear a 
physical motivation. }

This paper is organized as follows. In Section 2, we review key topics of entanglement spectral statistics, Clifford circuits, matchgate circuits, and give a formal definition of simulability. In Section 3, in direct relation to the work by Zhou et al.\ \cite{ShiyuChamon}, we show that a single SWAP gate inserted into a matchgate circuit is enough to generate a transition from Poisson to Wigner-Dyson level repulsion statistics in the entanglement spectra. In Section 4, we study the outcome of changing the input states, and show that there is a transition in entanglement complexity when input states are allowed to be products of arbitrary two-qubit states versus three-qubit states. In Section 5, we examine the effect of  \enquote*{conjugating} matchgate circuits with specific families of Clifford gates. We find that there exist families of quantum gates which produce Wigner-Dyson entanglement level statistics while remaining classically simulable. In this way, not only do we find the entanglement complexity of Clifford and matchgate hybrid circuits to be noticeably different from that of pure Clifford or pure matchgate circuits, but we also find that we can no longer identify Wigner-Dyson level statistics of the entanglement spectra to be a universal indicator for simulation complexity in a generic quantum circuit. {\color{black}As a consequence, the main result of this paper can be seen as analogous to previous work done by Lewenkopf \cite{PhysRevA.42.2431}, except within the context of entanglement spectra and quantum simulability. We therefore stress that the emergence of Wigner-Dyson distributed entanglement spectra in the absence of some universal gate element implies a richer understanding of quantum simulation complexity not seen in energy spectra.}



\section{Preliminaries}

\subsection{Entanglement Spectrum Statistics}

 Throughout this paper we will study the entanglement spectrum of a quantum state, defined as the eigenvalues of the reduced density matrix under an equal bi-partitioning of the system; i.e.,
\begin{align}
E_S= \{ p_k | p_k \in \text{ Spectra}(\rho_A)\}
\end{align}
for some pure state $|\psi\rangle$, and $\rho_A = Tr_B |\psi\rangle \langle \psi |$. In this work, we will focus on the level spacing of the entanglement spectra, captured by the ratio of adjacent gaps in the spectra, $r_k \equiv (\delta_{k-1})/(\delta_k)$, for $\delta_{k} = p_{k-1} - p_k$ and $p_k \geq p_{k+1}$. The level spacing indicates the way in which the eigenvalues are distributed, and allows us to use tools and comparisons from random matrix theory to better understand quantum states of interest \cite{Meh2004, Livan_2018}. \\
\indent We will deal with the spectra of two different level-spacing distributions; namely, Poisson and Wigner-Dyson. The Poisson distribution $P_{P}$ and Wigner-Dyson distribution $P_{WD}$ follow
\begin{subequations}
    \begin{align}
    P_{P}(r) &= \frac{1}{(1+r)^2} \\
    P_{WD}(r) &= \frac{(r+r^2)^{\beta}}{Z(1+r+r^2)^{1+3\beta /2}} 
\end{align}
\end{subequations}
 respectively, with $Z = \frac{4 \pi}{81 \sqrt{3}}$ and $\beta = 2$ \cite{Atas_2013,ShiyuChamon}. The Wigner-Dyson distribution for $\beta=2$ captures the distribution of spectra for the Gaussian unitary ensemble (GUE).

\indent Rather than working with the distribution itself, for our purposes it is often more instructive to work with a single numerical value that can characterize the full form of the distribution \cite{intfermions}. We introduce a modified ratio, defined as 
\begin{equation}
    \tilde{r}_k = \frac{\min ( \delta_{k}, \delta_{k+1} )}{\max ( \delta_{k}, \delta_{k+1} )}.
\end{equation} 
We note that the average over $k$ is $\tilde{r} \approx 0.386$ for the Poisson distribution, and $\tilde{r} \approx 0.603$ for the Wigner-Dyson distribution \cite{Atas_2013}. \\
\indent Besides studying the modified value $ \tilde r $, we also utilize the Kullback-Liebler divergence to study how the level-spacing distribution approaches either the Poisson or Wigner-Dyson distributions \cite{coeurjolly2006normalized}. The Kullback-Liebler divergence for some distributions $P_1(r)$ and $P_2(r)$ is defined as
\begin{equation}
    D_{KL} = \sum_i P_1(r_i) \log \left(P_1(r_i)/P_2(r_i) \right)
\end{equation}
and quantifies the difference between the distributions; i.e., how close $P_1(r)$ approximates the distribution $P_2(r)$. We will utilize both the modified ratio and the Kullback-Leibler divergence to study the entanglement spectra of matchgate circuits, comparing distributions to those known through the study of Clifford and Haar random states \cite{entspecirrev}. \\

\subsection{Clifford Circuits}
\indent The Clifford circuits are a highly studied class of simulable circuits. Their classical simulability comes from the celebrated Gottesman-Knill theorem. The theorem states that the dynamics of the generators of the stabilizer group for circuits composed of computational basis state inputs, Clifford gates, and intermediate adaptive measurement may be simulated in polynomial time \cite{nielsen00, gottesmanknill, CTab}. Access to arbitrary single qubit $Z$ rotation gates is enough to elevate the Cliffords to a universal gate set, though in practice it is often the $T$ gate that takes the Cliffords to universality. While the inclusion of $T$ gates or changes in the input state break Gottesman-Knill, the Cliffords can be used in other simulation tasks outside the applicability of Gottesman-Knill. \\
\indent We can consider a problem that is not simulable by Gottesman-Knill due to having non-computational basis state input. Let $U_{CL}$ be a $N$-qubit Clifford circuit with initial product state $| \Psi \rangle = | \psi_1 \rangle | \psi_2 \rangle ... | \psi_n \rangle $, with output Pauli $p = \mathcal{P}_1 \mathcal{P}_2 ... \mathcal{P}_n$ for $\mathcal{P}_i \in \{X, Y, Z, \mathcal{I} \}$. We can evaluate with the fact that $U_{CL}^{\dagger} p U_{CL} = p' = \mathcal{P}_1' \mathcal{P}_2' ... \mathcal{P}_n'$
\begin{equation}
    \langle \Psi | U_{CL}^{\dagger} p U_{CL} | \Psi \rangle = \prod_{i=1}^{n} \langle \psi_i | \mathcal{P}_i' | \psi_i \rangle
\label{eq:CSim}
\end{equation}
which can be evaluated in polynomial time. The distinction of types of simulation will be explored and clarified more in section \ref{sec:Sim}. 

Throughout this paper, when constructing random Clifford circuits, we construct a brickwork pattern of two-qubit Clifford gates, with each Clifford sampled following the work of Bravyi and Maslov \cite{Bravyi_2021}; i.e., a Hadamard-free Clifford, a layer of possible Hadamards, and finally another Hadamard-free Clifford. 

\subsection{Matchgates and Free Fermions} 
Matchgate circuits are a specific class of nearest-neighbor, parity preserving two-qubit gates $G(A,\,B)$ which may be written in the form \cite{Valiant,TerhalDivincenzo,JozsaMiyake, knill2001fermionic}
\begin{align}{\label{eq:MG}}
G(A,\,B)=\begin{pmatrix} a_{11} & 0 & 0 & a_{12} \\
0 & b_{11} & b_{12} & 0 \\
0 & b_{21} & b_{22} & 0 \\
a_{21} & 0 & 0 & a_{22} \end{pmatrix}.
\end{align}
We define
\begin{align}
A\equiv \begin{pmatrix} a_{11} & a_{12} \\ a_{21} & a_{22} \end{pmatrix},\quad B\equiv \begin{pmatrix} b_{11} & b_{12} \\ b_{21} & b_{22} \end{pmatrix}
\end{align}
and where we require that the matrices $A$ and $B$ (both in $U(2)$) have the same determinant (i.e., $\det(A)=\det(B)$) \cite{TerhalDivincenzo, JozsaMiyake}. Originally proven to be simulable in the context of the perfect matching problem on planar graphs \cite{Valiant}, small changes to the structure of matchgates such as relaxation of the determinant condition or the allowance of next-nearest neighbor gates \cite{Brod_2012, BrodXY, JozsaMiyake, Bravyi_2002} are sufficient to realize universal quantum computation. {\color{black} In terms of a single discrete gate to add to matchgates to recover universal computation, the SWAP gate is discussed most often. However, any two qubit parity preserving gate without the determinant condition will do.} Matchgates are simulable both on computational input states and on product state inputs \cite{TerhalDivincenzo, JozsaMiyake, Brod_2016}. \\\\
\indent  Although simulability in matchgate circuits can be understood via the connection to the perfect matching problem, a deeper physical intuition behind their simulability can be found from their connection to free fermions \cite{knill2001fermionic, TerhalDivincenzo}. Free fermion Hamiltonians are composed of quadratic fermionic creation and annihilation operators; i.e., operators $a_i, a_i^{\dagger}$ that obey the relations
\begin{subequations}
    \begin{align}
        \{a_i^{\dagger}, a_j^{\dagger} \} &= \{a_i, a_j \} = 0 \\
        \{a_i^{\dagger}, a_j \} &= \delta_{ij}
    \end{align}
\end{subequations}
where $\{\cdot\,,\,\cdot\}$ denotes the anti-commutator. The simulability of free fermion Hamiltonians can be derived from two key observations. The first is that conjugation of fermionic operators by unitaries defined by the exponentiation of free fermion Hamiltonians can be performed in polynomial scaling time and memory \cite{TerhalDivincenzo, JozsaMiyake}. The second is the fact that strings of fermionic operators can be contracted efficiently due to Wick's theorem \cite{TerhalDivincenzo, PhysRev.80.268}. 

When working with free fermions, it is convenient to work with the Hermitian Majorana operators, defined as
\begin{subequations}
    \begin{align}
    c_{2k-1} &= a_k + a_k^{\dagger} \\
    c_{2 k} &= i( a_k- a^{\dagger}_k ) \\
    \{c_i, c_j\} &= 2\delta_{ij}
    \end{align}
\end{subequations}
as the Majorana operators are often used in the diagonalization of free fermion Hamiltonians. They also impose a more useful anticommutation structure for fermion-to-qubit transformations \cite{Surace_2022, chien2020custom}. Fermion-to-qubit transformations are transformations that allow for the simulation of fermionic Hamiltonians on quantum computers by defining spin operators that obey the same anticommutation structure as the Majorana fermion operators. The most common fermion-to-qubit encoding is the non-local Jordan-Wigner transformation, defined as 
\begin{subequations}
    \begin{align}
        c_{2k-1} &= \left( \prod_{m=1}^{k-1} Z_m \right) X_k \\
        c_{2k-1} &= \left( \prod_{m=1}^{k-1} Z_m \right) Y_k
    \end{align}
\end{subequations}
\cite{1928ZPhy...47..631J}. While the action of conjugating the Majoranas with a Clifford is unclear, conjugating the equivalent spin operators with some Clifford defines a new set of anticommuting spin operators and a new fermion-to-qubit encoding. The qubit encoding of the Majorana fermions allows us to better examine hybrid Clifford and matchgate circuits. \\
\indent Matchgates specifically come from the exponentiation of two-qubit nearest neighbor free (quadratic) fermionic Hamiltonians, which may be written with Majorana fermions and real coefficients as
\begin{equation}
    \begin{aligned}
        H_{F} = i ( & \alpha_0 c_{2k-1} c_{2k+2} - \alpha_1 c_{2k}c_{2k+1} \\
         + & \beta_1 c_{2k-1}c_{2k+1} - \beta_2 c_{2k}c_{2k+2} \\
         - & \gamma_1 c_{2k-1}c_{2k} - \gamma_2 c_{2k+1}c_{2k+2}).
    \end{aligned}
\end{equation}
Via the Jordan-Wigner transform, we can define an equivalent spin Hamiltonian which produces matchgates, 
\begin{equation}
\begin{aligned}
    H_{S} = (&\alpha_0 Y_k Y_{k+1} + \alpha_1 X_k X_{k+1} \\
    + &\beta_1 Y_k X_{k+1} + \beta_2 X_k Y_{k+1} \\
    + &\gamma_1 Z_k + \gamma_2 Z_{k+1}).
\end{aligned}\label{eq:JWH}
\end{equation}
As shown in \cite{Jozsa_2009}, the exponentiation of {\it{any}} free fermion Hamiltonian can be written as the product of $O(n^3)$ matchgates. {\color{black} To make contact with universal computation in the language of fermions, we need fermionic interactions at nearest neighbor sites or beyond (or by including on-site interactions, if we allow for spin degrees of freedom)\cite{Bravyi_2002, Ji2019Nov}.} \\
\indent When constructing random matchgate circuits, we construct a brickwork of two-qubit matchgates. Each random matchgate is created by generating two random single qubit Haar unitaries (sampled from the Ginibre ensemble), fixing their determinants to be one, and then placing their terms in the two-qubit unitary following equation \ref{eq:MG}. 

\subsection{Simulability}{\label{sec:Sim}}

Attempts to formalize the treatment of simulability are often difficult due to the nature of computational complexity. \cite{watrous2008quantum, Whitfield_2013}. Nevertheless, we can still use results in complexity theory to highlight exactly what it means to be simulable or not, and define the complexity classes we will consider throughout this work. We define 
\begin{itemize}
    \item $\textsc{\textbf{P}}$: The class of decision problems solvable on a deterministic Turing machine in polynomial time.
    \item $\textsc{\textbf{NP}}$: The class of decision problems solvable on a nondeterministic Turing machine in polynomial time. It is equivalently the class of problems verifiable by a deterministic Turing machine in polynomial time. 
    \item $\textsc{\textbf{BQP}}$: The class of decision problems solvable on a quantum computer in polynomial time, with a bounded probability of error using a polynomial sized quantum circuit \cite{nielsen00}.
    \item $\textsc{\textbf{\#P}}$: The class of counting problems of the number of accepting paths in a non-deterministic Turing machine. $\textsc{\textbf{\#P}}$ (and  $\textsc{\textbf{NP}}$) are both believed to contain problems that are not in $\textsc{\textbf{BQP}}$ \cite{aaronson2009bqp, Bennett_1997}.
\end{itemize}

We will pull heavily from the definitions used in papers about Clifford and matchgate complexity with varied supplemental resources \cite{JoszaCliffC, MatchwOutside}. We define \enquote*{simulability} as the ability to perform a task associated to a quantum computer in polynomial time on a classical machine. A \enquote*{task} in this case will be associated with the outputs of a certain circuit with the five different attributes, namely
\begin{enumerate}
    \item The family of gates $U_k$. 
    \item The allowed input states; e.g., \textsc{\textsc{IN(Bits)}} being a computational basis state, or \textsc{IN(Product)} being a product state.
    \item The types of intermediate measurement allowed; e.g., non-adaptive (\textsc{Nonadapt}, which is considered to be equivalent to unitary circuits) or adaptive measurements (\textsc{Adapt}). Throughout this paper we never consider intermediate measurement, so we are always in the \textsc{Nonadapt} case.
    \item The type of output; e.g., OUT(1) or \textsc{OUT(Many)}. OUT(1) corresponds to the output distribution of only a single qubit, while \textsc{OUT(Many)} corresponds to the distribution of multi-qubit outputs.
    \item The type of simulation; i.e., strong or weak simulation \cite{JoszaCliffC}. \textit{Weak Simulation} is the task of sampling from the output distribution $p(y_{i_1}, y_{i_2}...y_{i_m})$ over $m$ outputs for $y_i = \{0,1 \}$. This kind of simulation best resembles the process of performing a task on a quantum computer: a task in which after each circuit iteration, the output is a sample from the probability distribution of the circuit. In contrast, {\it{strong simulation}} is where all marginal probabilities for the circuit can be computed in polynomial time; i.e., we have efficient computation of any desired output probability. For consistency, through the rest of this work, we will label simulation types as \textsc{Strong} and \textsc{\textsc{Weak}}. 
\end{enumerate}
The Gottesman-Knill theorem tells us that the task of Clifford gates, \textsc{Adapt}, \textsc{IN(Bits)}, \textsc{OUT(Many)}, and \textsc{\textsc{Weak}}, is simulable \cite{gottesmanknill, nielsen00, JoszaCliffC}, or that we can efficiently sample from the probabilities of multiple outputs of Clifford circuits in polynomial time. Equation \ref{eq:CSim}, with choice of Pauli $p=Z$, tells us that the task of Clifford gates, \textsc{Nonadapt}, \textsc{IN(Prod)}, OUT(1), and \textsc{Strong} is classically simulable, as  $p_0 - p_1 = \langle \psi | U^{\dagger}ZU | \psi \rangle$, and with the knowledge $p_0 + p_1 = 1$, getting the probabilities of both measuring $0$ or $1$ is solvable. This example highlights the fine line between simulability and non-simulability, as the task of Clifford gates, \textsc{Nonadapt}, \textsc{IN(Prod)}, \textsc{OUT(Many)}, and \textsc{Strong} is $\textsc{\textbf{\#P}}$ hard \cite{JoszaCliffC}. \\
\indent Matchgates present a comparison point between the Cliffords not only because they are simulable for different reasons, but because they have different complexities for different tasks. As an example, while the task of Clifford gates, \textsc{Adapt}, \textsc{IN(Prod)}, \textsc{OUT(Many)}, and \textsc{Weak} represents a task that can perform universal quantum computation (and, thus, is not simulable), for matchgates, one finds that \textsc{Adapt}, \textsc{IN(Prod)}, \textsc{OUT(Many)}, and \textsc{Weak} is classically simulable, and remains so even if we desire \textsc{Strong} simulation (in the Clifford case changing to \textsc{Strong} simulation makes simulation $\textsc{\textbf{\#P}}$ hard \cite{JoszaCliffC}). \\
\indent The differences in simulation complexity, and the wide range of efficient simulations with matchgate circuits, motivate us to study their entanglement complexity as a comparison to Clifford and Haar random circuits. By studying circuits with separate simulability conditions, we can gain a better understanding of the relationship between entanglement spectrum and simulability.

\section{Matchgates With Injected SWAPs}
\label{sec:SWAPs}

\subsection{Setup}

\begin{figure}
\begin{center}
\begin{adjustbox}{width=0.45\textwidth}
    \begin{quantikz}
        &\gate{R}&\gate[2]{M}& & \ \ldots\ & \swap{1} & &\gate[2]{M}& & \ \ldots\ \\
        &\gate{R}& &\gate[2]{M} & \ \ldots\  & \targX{} & \swap{1} & &\gate[2]{M} & \ \ldots\ \\
        &\gate{R}&\gate[2]{M}& & \ \ldots\  & &\targX{} &\gate[2]{M}& & \ \ldots\ \\
        &\gate{R}& &\gate[2]{M} & \ \ldots\  & & & &\gate[2]{M} & \ \ldots\ \\
        &\gate{R}&\gate[2]{M}& & \ \ldots\  & \swap{1} & &\gate[2]{M}& & \ \ldots\ \\
        &\gate{R}& & &\ \ldots\  &\targX{} & & & &\ \ldots\ \\
    \end{quantikz}
\end{adjustbox}
\end{center}
\caption{Setup of circuits with injected SWAP gates. Starting with single qubit real rotations to make arbitrary product states ($R$), layers of matchgates are applied ($M$), before some number of SWAP operations are performed (linked $x$ markers). After SWAP injection, matchgates further mix the system. }
\label{fig:Setup}
\end{figure}
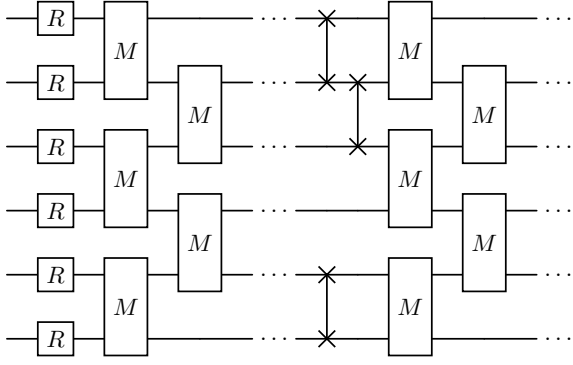

We numerically simulate a number of random matchgate circuits for circuits of qubit sizes $N$ = 12, 14, 16, 18, and 20. As seen in Fig.\ \ref{fig:Setup}, we begin with generating a random product state, through applying a sequence of random single qubit real rotations on the $|00...0\rangle$ state. After initializing, we evolve under a brickwork circuit of $N^2$ layers. Even layers consist of matchgates acting on nearest neighbor even pairs $\{ (0,1),(2,3),...,(N-2, N-1) \}$, while odd layers have gates acting on odd pairs $\{ (1,2),(3,4),...,(N-3, N-2) \}$. After the quadratic layers of matchgates, a number of swapping operations SWAP are applied. The number of SWAPs is varied from $1$ to $N-1$, applied in a maximum of two layers. SWAPs over all even pairs are applied first in one layer, followed by a second layer where SWAPs over odd pairs are applied. After the SWAPs are injected into the circuit, some number of layers of matchgates ($100$ for $N$=12, $120$ for $N$=14, $140$ for $N$=16, $160$ for $N$=16, and $180$ for $N$=20) are applied corresponding to the circuit sizes given above. The number of circuits generated depends on the system size (1050 for $N$=12, $N$=14, and $N$=16, 525 for $N$=18, and 225 for $N$=20).

\subsection{Numerical Results for $\langle \tilde r \rangle$}

\begin{figure}
    \hspace{-6mm}
    \includegraphics[width=1.1\linewidth]{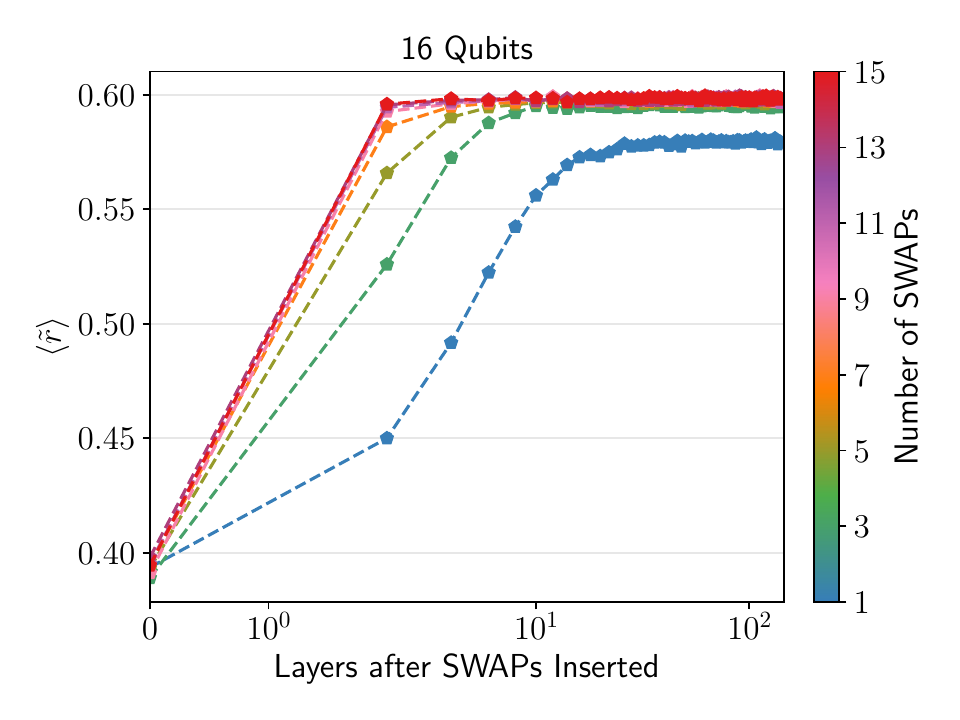}
    \caption{$\langle \tilde r \rangle$ as a function of layers after SWAP injection in 16 qubit circuits. For a single SWAP gate, the rate of growth towards $\langle \tilde r \rangle$ that characterizes the Wigner-Dyson distribution is slower. For a finite size system, a single SWAP is not enough to necessarily converge exactly to the Wigner-Dyson distribution.}
    \label{fig:16SWAP}
\end{figure}

\begin{figure}
    \hspace{-6mm}
    \includegraphics[width=1.1\linewidth]{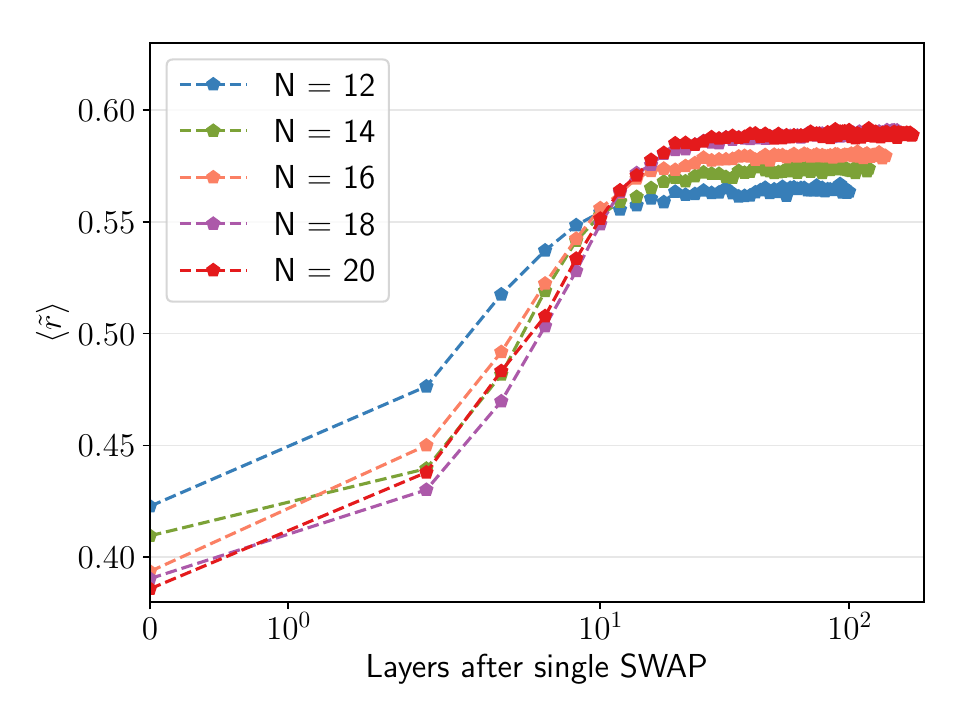}
    \caption{$\langle \tilde r \rangle$ as a function of layers after a single SWAP insertion for various circuit sizes. We see that as we increase $N$, the $\langle \tilde r \rangle$ gets closer to the accepted value for Wigner-Dyson distributions.}
    \label{fig:1SWAPN}
\end{figure}

\indent In Fig.\ \ref{fig:16SWAP} we study $\langle \tilde r \rangle$, the average of $\tilde r$ over all iterations, for a system size of $16$ qubits, observing behavior as a function of layers post SWAP injection. We notice that, for a finite-sized system and for a small number of injected SWAPs, we don't quite approach the Wigner-Dyson distribution, but instead interpolate close to it as the number of SWAPs injected increases. This motivates us to study the relationship between $\langle \tilde r \rangle$ and the number of injected SWAPs as we change system size. \\
\indent In Fig.\ \ref{fig:1SWAPN}, we note that in the presence of a fixed number of injected SWAPs, as we increase system size we interpolate closer to the $\langle \tilde r \rangle$ value for the Wigner-Dyson distribution. We also note a detail specifically about the $\langle \tilde r \rangle$ value at $0$ layers post-SWAP insertion, or, equivalently, the layer before the SWAP is inserted. Note that, while for small Clifford circuits pre-$T$ gate insertion, the value of $\langle \tilde r \rangle$ is at times well below the accepted Poisson value of $0.39$ \cite{ShiyuChamon}, the analogous scenario for matchgates under SWAP insertion is reversed. For small systems before SWAP injection, $\langle \tilde r \rangle$ exceeds the Poisson value, and converges towards the Poisson value as system size increases. \\

\indent We now come to the main focus of this section, studying $\langle \tilde r \rangle$ in the infinite time limit (i.e., at the point in which further evolution under matchgates after SWAP insertion has no large change to the system). Based on our findings above, the value of $\langle \tilde r \rangle_{\infty}$ is dependent both on the size of the system and on the number of SWAPs injected into the circuit. In this section we will look at $\langle \tilde r \rangle$ for only a single SWAP inserted, and look at behavior with respect to only system size. This is in contrast to the analysis done by Zhou et al.\ \cite{ShiyuChamon} for the Cliffords plus $T$. Zhou et al.\ are able to fit $\langle \tilde r \rangle_{\infty}$ and the deviations from the accepted Wigner-Dyson value to $N \cdot N_{T}$, where $N_T$ is the number of $T$ gates. In Appendix \ref{app:ClifC}, we will show analysis done on fitting to $N\cdot N_{\textrm{SWAP}}$, and discuss the reasons for different behavior than seen in the Clifford plus $T$ case. \\
\indent Error on $\langle \tilde r \rangle_{\infty} $ is calculated as follows: for each randomly generated circuit, we define $\tilde r_{\infty} $ as the average $ \tilde r$ value over the final 40 layers of the circuit to guarantee calculating the average only over the infinite time regime (convergence to infinite time regime takes maximum $O(n)$ time as the effect of the injected gate spreads through the circuit). For each circuit ran, the $\tilde r_{\infty} $ value is saved. The final recorded data of $\langle \tilde r \rangle_{\infty} $ is recorded as the average of all saved $\tilde r_{\infty} $ values, and the error bars are generated from the standard deviation of all $\tilde r_{\infty} $ saved. 
\\

\begin{figure}
    \hspace{-6mm}
    \includegraphics[width=1.1\linewidth]{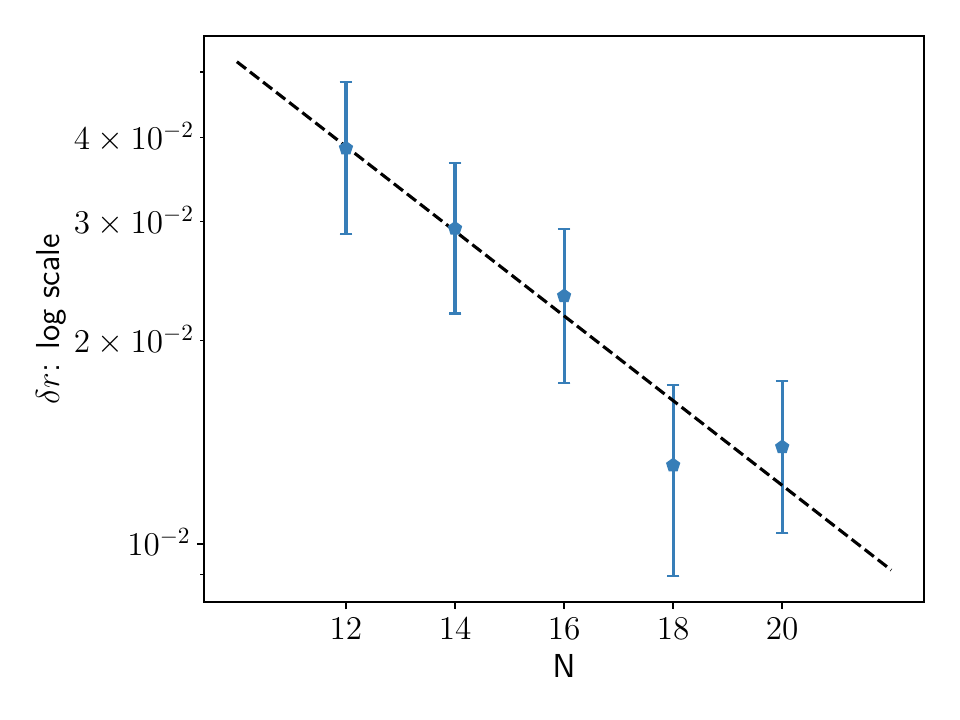}
    \caption{$\delta r$, the deviation from the calculated $\langle \tilde r \rangle_{\infty}$ to the Wigner-Dyson value of $\langle \tilde r \rangle \approx 0.603$ on a log scale, fitted to $\delta r = r_0 e^{-\gamma N}$. For a single SWAP gate, we find a roughly linear fit of the log of $\delta r$, showing that in the limit of large $N$ we expect deviations going to zero.}
    \label{fig:Deviations}
\end{figure}

\indent We now look at the infinite time limit for a single SWAP inserted; specifically, we look at fitting the deviations from the calculated $\langle \tilde r \rangle_{\infty}$ to the accepted large system size value for GUE distributed matrices of $\langle \tilde r \rangle \approx 0.603$. In Fig.\ \ref{fig:Deviations} we plot the log of deviations, fit to $\delta r = r_0 e^{-\gamma N}$ for fitting parameters $r_0$ and $\gamma$ and find roughly a linear relationship. As deviations $\delta r$ go to zero as $N$ tends towards infinity, we find strong evidence that a single SWAP in the thermodynamic limit is enough to transition from Poisson entanglement level statistics to Wigner-Dyson. \\
\indent {\color{black} We may also connect the SWAP directly back to fermionic systems. It is known that for a general parity preserving non-matchgate element like SWAP that there exists a gate decomposition 
\begin{equation}
    U_{PP} = (RZ_1 \times RZ_2) e^{i(a XX + bYY + cZZ)}(RZ_3 \times RZ_4)
\end{equation} 
while for matchgates, the decomposition is 
\begin{equation}
    U_{PP} = (RZ_1 \times RZ_2) e^{i(a XX + bYY)}(RZ_3 \times RZ_4)
\end{equation} 
\cite{Brod_2011, mocherla2024extending}.
Using the Jordan Wigner mapping and observing that $Z_iZ_{i+1} = c_{2i}c_{2i+1}c_{2i+2}c_{2i+3}$ it is clear that to make a SWAP gate requires a fermionic interaction. 
In other words, the injection of a SWAP gate can be thought of as adding a single instance of interaction in a free fermion system. Therefore, the conclusion that a single SWAP is enough to induce Wigner-Dyson distributed entanglement spectra is equivalent to saying that a
fermionic interaction is sufficient for 
changing entanglement spectra statistics.} 

\subsection{Entanglement Entropy vs. Entanglement Complexity}

There is another important distinction we need to draw compared to the Clifford plus $T$ case: the disconnect between entanglement complexity and the amount of entanglement. In Clifford circuits on product state inputs, a single injected $T$ gate is enough to induce Wigner-Dyson entanglement level statistics, but gives no change to the amount of entanglement entropy. This is due to the fact that Clifford circuits, on average, generate the same amount of entanglement as Haar random circuits \cite{PhysRevLett.71.1291, ShiyuChamon}. We show that for matchgates with injected SWAPs, this is not the case: not only do matchgates not generate the Haar average amount of entanglement, but after a single SWAP gate injected in the thermodynamic limit as $N\rightarrow \infty$, though entanglement complexity sharply changes, the entanglement entropy does not approach the Page entropy.  The Page entropy is the average entropy of Haar random states, defined as, for Hilbert space dimension $mn$ and $m \leq n$ \cite{PhysRevLett.71.1291},
\begin{equation}
    S_{m,n} = \sum_{k=n+1}^{mn}\frac{1}{k} - \frac{m-1}{2n}.
\end{equation} \\

\begin{figure}
    \hspace{-6mm}
    \includegraphics[width=1.1\linewidth]{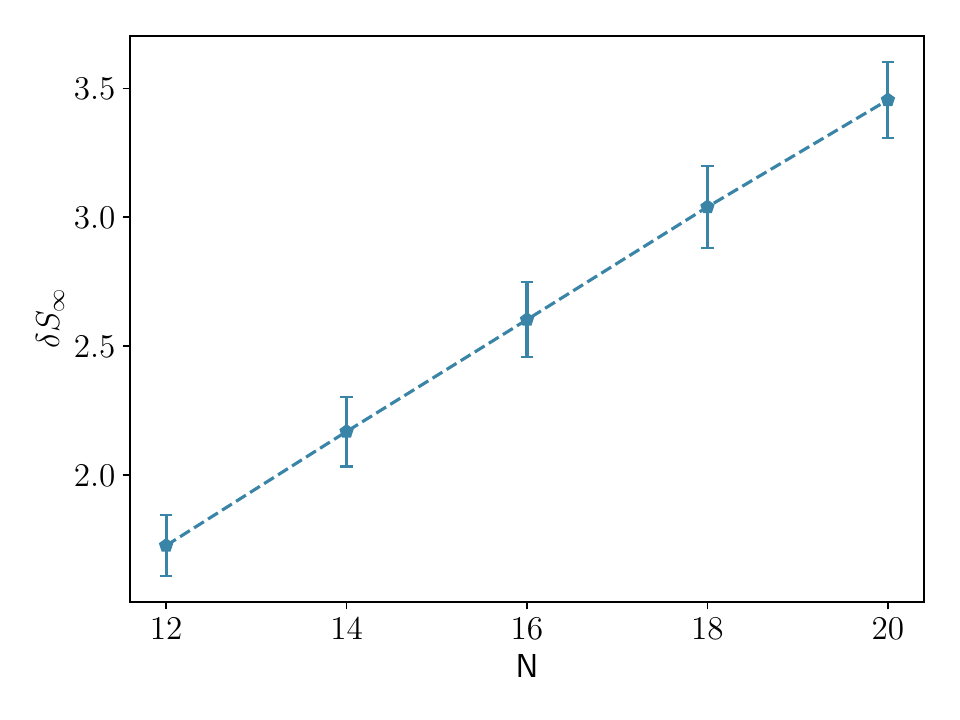}
    \caption{For a single SWAP inserted, the deviation from the Page entropy increases linearly as a function of system size, {\color{black} in the infinite time regime after insertion.} }
    \label{fig:1SWAPent}
\end{figure}

\indent We are interested in the deviation in the matchgate entropy to the Page entropy with a single SWAP inserted. {\color{black} We examine inserting more than one SWAP in Appendix \ref{App:Ent}.} As seen in Fig.\ \ref{fig:1SWAPent}. For only a single SWAP inserted, the deviation from the average entropy in a matchgate plus SWAP circuit compared to that of a Haar circuit increases linearly as a function of $N$. Though the states we can reach through these matchgate plus SWAP circuits are complex in entanglement structure, they do not have the same average entropy as Haar states do.\\ {\color{black}\indent As such, we conclude that there is no connection between entanglement spectra and the saturation of entropy to the Page limit. More specifically, the fact that matchgates do not have the same entanglement entropy as the Page limit implies that matchgates cannot form a $2$-design (see section 3.4.1 of \cite{Liu_2018}). Thus, entanglement spectra of some state is not connected to how well the unitaries which evolved said state approximate distributions over the Haar measure.} \\

\indent This highlights the divide between entanglement complexity and the amount of entanglement as quantified by entanglement entropy. As we see in matchgate plus SWAP circuits, we can have complex entanglement that is not necessarily from states near the Page entropy.

\section{Matchgates with Various Input States}
\label{sec:Input}
We now come to the second part of our study. In the context of matchgate circuits, studying how varying input states changes the complexity of entanglement. We have already discussed above the simulation of matchgate circuits on product state inputs. Extending to states which are products of two qubit entangled inputs, without any adaptive measurements, remains \textsc{Strong} classically simulable on OUT(\textsc{Many}) qubit outputs \cite{MatchwOutside}. On products of $O(1)$ qubit entangled state inputs, the simulation of OUT(1) is \textsc{Strong}, while for OUT(\textsc{Many}) the same simulation is  $\textsc{\textbf{\#P}}$ hard \cite{MatchwOutside}.\\ 

\begin{figure}
    \hspace{-6mm}
    \includegraphics[width=1.1\linewidth]{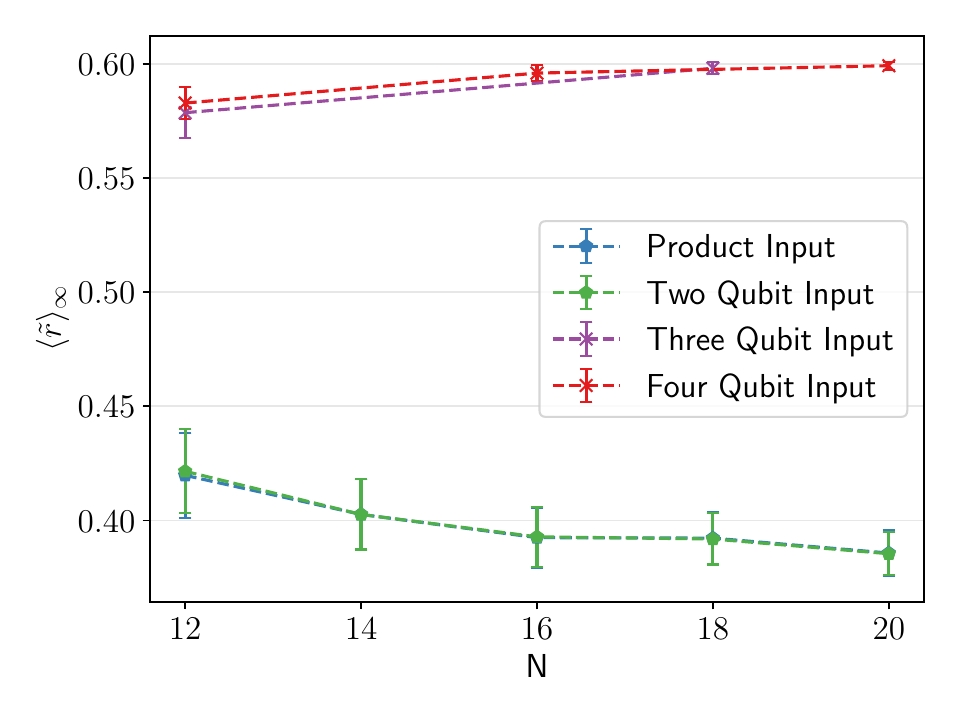}
    \caption{Plot of $\langle \tilde r \rangle_{\infty}$ versus circuit size, with different input states. While product and two-qubit input states have a near perfect overlap, for three and four-qubits the complexity of entanglement jumps to the Wigner-Dyson value. There is a system size dependence, as for smaller system sizes for three and four-qubit inputs the $\langle \tilde r \rangle_{\infty}$ value isn't quite the GUE value, but for larger system sizes it approaches it.} 
    \label{fig:rInjSWAP}
\end{figure}

\indent We find a sharp jump in the complexity of entanglement when going from two- to three-qubit entangled state inputs. More specifically, we find that both the entanglement and entanglement spectrum are identical for one and two-qubit state inputs. For three and four-qubit entangled state inputs, though entanglement entropy does not go to the average Page entropy, the entanglement spectrum approaches a Wigner-Dyson distribution. \\

\indent We see in Fig.\ \ref{fig:rInjSWAP} that the $\langle \tilde r \rangle_{\infty}$ values for different system sizes are nearly identical for product and two-qubit state inputs, starting from above the Poisson value of $ \langle \tilde r \rangle_{\infty} \approx 0.39$ for small system sizes then slowly approaching it as size increases. For input sizes of three and four qubits, we see a sharp jump to the Wigner-Dyson value of $ \langle \tilde r \rangle_{\infty} \approx 0.60$, with small deviations for small circuit sizes. \\

\begin{figure}
    \hspace{-6mm}
    \includegraphics[width=1.1\linewidth]{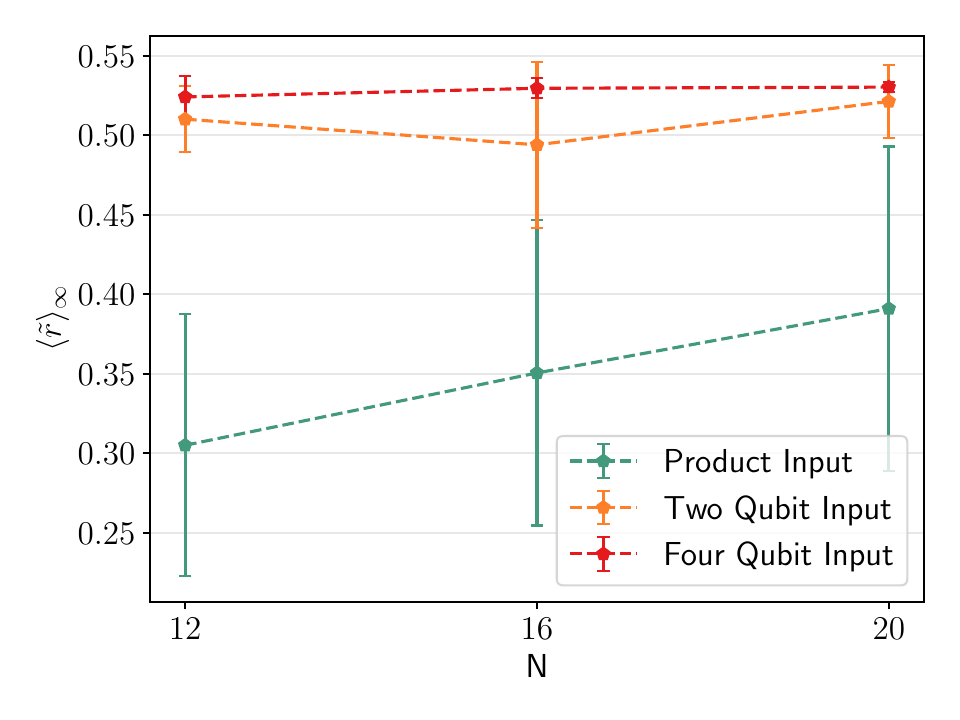}
    \caption{{\color{black} Plot of $\langle \tilde r \rangle_{\infty}$ versus circuit size, with different input states, but for the Cliffords. We see different behavior as compared to matchgates; rather than sharp jump with respect to certain input states, there is more smooth interpolation between between different distributions of entanglement spectra. }} 
    \label{fig:rinputcomp}
\end{figure}

\indent As matchgates on product state and matchgates on products of two-qubit entangled inputs can both be made from matchgate circuits with an ancilla in the $|+\rangle$ state, we do not expect any difference between the two \cite{MatchwOutside, Brod_2016}. While matchgates on three or four-qubit entangled inputs are technically simulable (see Theorem 1 of Hebenstreit et. al \cite{MatchwOutside}), they nonetheless have Wigner-Dyson entanglement spectra. This provides further evidence that there is no device for building arbitrary states more than two qubits with matchgates and a small number of resources like the $|+\rangle$ state. To build arbitrary three and four-qubit entangled states, universal resources such as a SWAP must be used. This is in contrast to the Clifford case with varied input where there is no sharp jump, as seen in Fig.\ \ref{fig:rinputcomp}. {\color{black} The difference in the two cases can be understood by recognizing that input states are different resources for Cliffords and matchgates \cite{JoszaCliffC, MatchwOutside}}. \\
\indent {\color{black} Connecting input states back to fermionic systems is a non-trivial task, unlike connecting non-matchgate elements back to a fermionic interaction. One simple example we may use to connect back to fermionic models is the product state input. The ability to have an ancilla in the $|+\rangle$ state can be understood by including number and parity non-conserving terms in the electronic Hamiltonian. Consider the Hamiltonian 
\begin{equation}
    H_{GS} = \sum_{u,v}h_{uv}c_u c_v + \sum_u s_u c_u
\end{equation}\\
for which the time dynamics $U^{\dagger}c_u U$ with $U = e^{iH_{GS}}$ can be efficiently simulated (as with the purely quadratic case), except with summations over $2n+1$ instead of $2n$ \cite{jozsa2015jordanwigner, henderson2024hartreefockbogoliubov}. By virtue of a Majorana representation, the linear fermionic term $c_1$ maps to $X$, thus allowing for arbitrary $X$ rotations on the a single qubit. This is enough to implement Hadamard, and thus arbitrary product states \cite{Brod_2016}. \\
\indent To build input states of products of two, three, and four qubit entangled states requires resources beyond the addition of linear fermionic terms. Therefore, we note that some arbitrary fermionic interaction we add to our circuit 
need not always lead to a change in entanglement spectra. It is not only the presence of fermionic interaction which causes a change in spectra, but rather it matters where and when the interacting fermionic gate impacts time evolution. }

\indent In summary, looking at varied input states, there is a sharp jump in the complexity of entanglement when input states go from being the product of two-qubit entangled states to being the product of three-qubit entangled states, respectively. We discuss more observations around the average entropy of these systems in Appendix \ref{App:Ent}. 

\section{Clifford-conjugated Matchgates}\label{sec:Conj}

In Jozsa and Miyake's paper on \textsc{IN(Prod)} and OUT(1) matchgate simulation, the authors discuss the case of Clifford-conjugated matchgate circuits \cite{JozsaMiyake}. For a set of algebraic operators $c_{\mu}$ which obey a Clifford algebra (in this case, the Majorana operators), conjugation by some unitary does not change their algebraic properties. We recall that matchgates can be defined as the exponentiation of a sum of Pauli terms via the Jordan-Wigner transformation, as seen in equation \ref{eq:JWH}. We find that by conjugating these operators by Cliffords we retain algebraic structure and that quadratic products remain quadratic. This is equivalent to the mapping from one fermion-to-qubit encoding (Jordan-Wigner) to some other (but still valid) encoding. This allows for the simulation of expectation value outputs, and thus the efficient simulation of Clifford and matchgate hybrid circuits. We define Clifford circuit $C$, matchgate circuit $U_{MG}$, \textsc{IN(Prod)}, OUT(1), and \textsc{Strong} simulation, via
\begin{equation}
    \langle \Psi_{prod}|C^{\dagger}U_{MG}^{\dagger}CZ_kC^{\dagger}U_{MG}C| \Psi_{prod} \rangle 
\end{equation}

\noindent so long as $Z_k$ can be written as a product of a constant scaling number of conjugated encoded operators. The simulation cost is explicitly $O(d)$ for 
\begin{equation}
    Z_k = \prod_{i=1}^{d} c_i
\end{equation}
So long as $d$ is not $O(n)$, expectation value outputs can be simulated in polynomial time \cite{JozsaMiyake}. 

 Though the new simulation problem of the expectation value above can be seen as conjugation of the matchgate circuit, we can define an equivalent problem of simulating $U_{MG}$ on a set of inputs $C|\Psi_{prod}\rangle$ with expectation value output $CZ_kC^{\dagger}$, which we will define as \enquote*{conjugation}. \\
\indent We focus on \enquote*{conjugation} by two different circuits, both presented in Jozsa and Miyake \cite{JozsaMiyake}. The two circuits are (gates on the left acting first)
\begin{subequations}
\begin{align}
    \begin{aligned}
            { C_1} &= (CNOT_{1,2}CNOT_{2,3}...CNOT_{N-1,N} \\
            &H_1 H_2 ... H_N) \text { and }
    \end{aligned} \\
    \begin{aligned}
        { C_2} &= (CNOT_{1,2}CNOT_{3,4}...CNOT_{N-2,N-1} \\
        & CNOT_{3,2}CNOT_{5,4}...CNOT_{N,N-1}).
    \end{aligned}
\end{align}
\end{subequations}

\begin{figure}
   \hspace{-6mm}
    \includegraphics[width=1.1\linewidth]{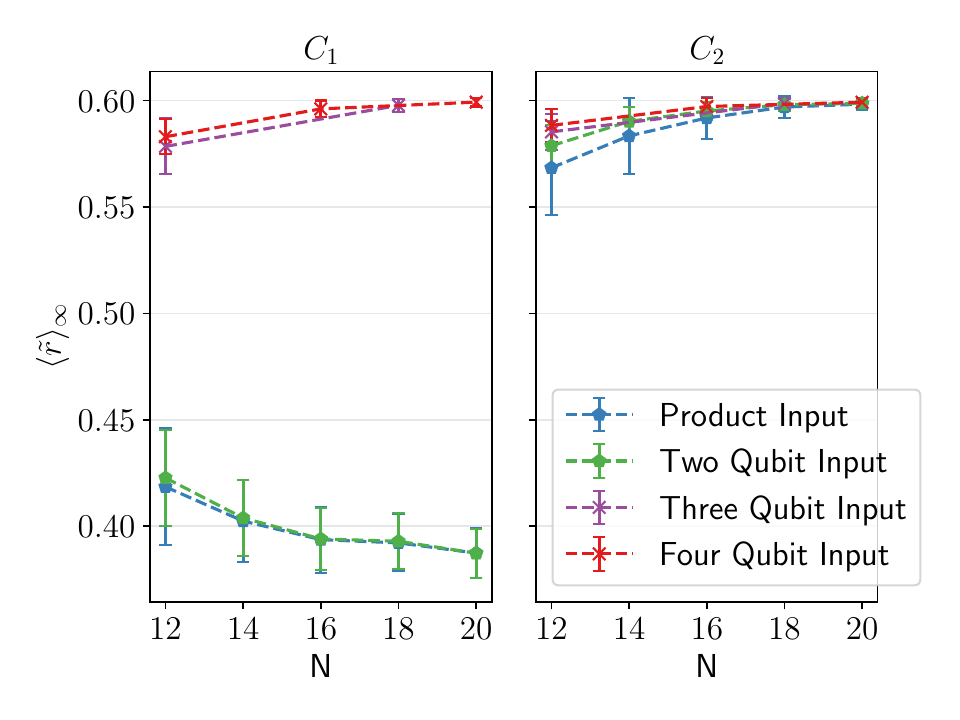}
    \caption{$\langle \tilde r \rangle_{\infty}$ for matchgate circuits on various input states with a various input states followed by Clifford $C_1$ and $C_2$. Entanglement spectrum statistics for $C_1$ match the same behavior as they do for matchgates on input states without the Clifford; for product and two-qubit entangled inputs we have Poisson distributed entanglement spectra, with a jump in three and four qubit entangled state inputs. For $C_2$, even on product states, we interpolate to Wigner-Dyson as a function of system size.} 
    \label{fig:t1C}
\end{figure}

\noindent \enquote*{Conjugation} with $C_1$ on a matchgate circuit does not change entanglement level statistics as compared to matchgates without conjugation, as seen in Fig.\ \ref{fig:t1C}. On single qubit and two-qubit input states entanglement spectrum statistics are Poisson, while for three and four-qubit inputs there is an immediate jump to Wigner-Dyson. \\

\indent When \enquote*{conjugating} with $C_2$, as seen in Fig.\ \ref{fig:t1C}, the entanglement spectrum interpolates to Wigner-Dyson as system size increases, irrespective of the class of input states. Thus, though expectation value $\langle \Psi_{prod}|C_2^{\dagger}U_{MG}^{\dagger}C_2Z_kC_2^{\dagger}U_{MG}C_2| \Psi_{prod} \rangle $ is simulable (albeit in $O(n^6)$ time, see \cite{JozsaMiyake}), circuits of the form $U_{MG}C_2| \Psi_{prod} \rangle$ produce states with complex entanglement. \\
\indent {\color{black} This result can by interpreted physically, by understanding what fermionic system Clifford-conjugated matchgate circuits correspond to. At the level of a spin chain, matchgates correspond to the time dynamics of a spin Hamiltonian that can be mapped back to some free fermion Hamiltonian. Conjugating matchgates with Cliffords results in the time dynamics of a conjugated spin Hamiltonian that can still be mapped to free fermions \cite{Chapman_2020, chapman2023unified}. Thus, we can change the entanglement spectra of some fermionic circuit just by changing the spin system which represents the fermionic Hamiltonian.} \\ 
\indent To the present authors' knowledge, the Clifford/matchgate hybrid circuit is the {\it first example} of a quantum circuit which is composed of simulable gate set elements yet has Wigner-Dyson distributed entanglement spectrum statistics, without any perturbing universal element. Note that, although quantum automata circuits (as introduced in \cite{PhysRevB.100.214301}) may produce a complex entanglement spectrum, prior work fails to show if the classical Monte Carlo method used converges in polynomial runtime up to digits of precision $d$ \cite{PhysRevB.100.214301, Iaconis_2021}. As a consequence, although quantities of physical interest in quantum automata circuits may be calculated on a classical computer, the explicit conditions for simulability of quantum automata circuits in $\textsc{\textbf{P}}$ are, as of yet, unknown. 

\section{Conclusion}
In this paper, we have discussed the level statistics of entanglement spectra in matchgate circuits with additional resources in the form of injected SWAP gates, varied input state sizes, and with conjugation by Clifford gates. For matchgates with injected SWAPs, we found that, much as in the case of the Cliffords with a single $T$ gate, a single SWAP is enough to produce Wigner-Dyson level statistics in the entanglement spectrum as the size of the circuit tends towards infinity. We note, however, that a single SWAP is not enough to take matchgate circuits to the Page entropy  \cite{PhysRevLett.71.1291}. Thus, while a single SWAP can randomize matchgate circuits, it is not enough to randomize matchgate circuits over the entire Hilbert space. \\
\indent We also show that, as we go from two- to three-qubit entangled inputs, we find a jump in complexity from Poisson to Wigner-Dyson level statistics in the entanglement spectrum. This further highlights that while there are constructions that can produce arbitrary two-qubit inputs for matchgate circuits, there cannot be a construction for arbitrary three-qubit states without use of the SWAP or other non-matchgate gate. \\
\indent Finally, we show that there exist circuits with \textsc{IN(Prod)}, OUT(1), and \textsc{Strong} which are simulable, while having Wigner-Dyson distributed entanglement spectra. This presents a shift from the idea that the entanglement spectra can always be used as an indicator for simulability, or when a problem is \enquote*{hard}.\\

\indent {\color{black} } \\
\indent Overall, we provide examples clarifying the relationship between simulability, entanglement entropy, and entanglement spectra, showing that all three can be independent of one another. There are simulable problems that have non-area law but non-Page entropy which have Poisson entanglement spectra, and others that have Wigner-Dyson entanglement spectra. Just because a given circuit is classically simulable does not guarantee that this circuit will exhibit maximal entropy, nor does that mean the entanglement spectra follows a Poisson distribution. We {\color{black} introduce a BGS-like} conjecture {\color{black} for the relationship between entanglement spectra statistics and complexity}: that there exists no problem with no simulable aspects that has Poisson level statistics in the entanglement spectrum. {\color{black} We similarly conjecture that} having Wigner-Dyson level statistics in the entanglement spectra is a necessary {\it but not sufficient} condition for a problem to be non-simulable.  \\
\indent This work also raises open questions about both matchgates and entanglement spectrum statistics. For the latter, one might ask if there is an analytical approach which proves some simulable circuits have Poisson entanglement level statistics while others exhibit Wigner-Dyson? Understanding this line of thought would reveal more about the nature of simulable problems, and will be a topic of future exploration. For matchgates, while the same on one and two-qubit inputs, when looking at entanglement metrics matchgates on one and two qubit inputs are not drastically different than on three and four-qubit entangled inputs. Future work will involve studying the difference between matchgate circuits with varied inputs, with the hope to find analytical tools to rigorously quantify the distance between the outputs of different circuits. 

\section*{Acknowledgements}
The authors thank Xiao Chen, Andrew Cupo, Joseph Gibson, Brent Harrison, and Stefanos Kourtis for discussion and suggestions on various aspects of the paper. {JDW and AMP are supported by the Office of Science, Office of Advanced Scientific Computing Research under programs Fundamental Algorithmic Research for Quantum Computing and Optimization, Verification, and Engineered Reliability of Quantum Computers project. JDW holds concurrent appointments at Dartmouth College and as an Amazon Visiting Academic. This paper describes work performed at Dartmouth College and is not associated with Amazon.}

\section*{Code Availability}
Code used in this project is available upon reasonable request {by contacting the corresponding author.}

\bibliographystyle{quantum}
\bibliography{Pbib.bib}

\begin{thebibliography}{10}

\bibitem{DOIKOU_2010}
Anastasia Doikou, Stefano Evangelisti, Giovanni Feverati, and Nikos Karaiskos.
\newblock ``{Introduction} {to} {Quantum} {Integrability}''.
\newblock \href{https://dx.doi.org/10.1142/s0217751x10049803}{International
  Journal of Modern Physics A {\bf 25}, 3307--3351}~(2010).

\bibitem{Scaramazza_2016}
Jasen~A. Scaramazza, B.~Sriram Shastry, and Emil~A. Yuzbashyan.
\newblock ``Integrable matrix theory: Level statistics''.
\newblock \href{https://dx.doi.org/10.1103/PhysRevE.94.032106}{Phys. Rev. E
  {\bf 94}, 032106}~(2016).

\bibitem{Gubin_2012}
Aviva Gubin and Lea~F. Santos.
\newblock ``Quantum chaos: An introduction via chains of interacting spins
  1/2''.
\newblock \href{https://dx.doi.org/10.1119/1.3671068}{American Journal of
  Physics {\bf 80}, 246--251}~(2012).

\bibitem{Rabson_2004}
D.~A. Rabson, B.~N. Narozhny, and A.~J. Millis.
\newblock ``Crossover from poisson to wigner-dyson level statistics in spin
  chains with integrability breaking''.
\newblock \href{https://dx.doi.org/10.1103/PhysRevB.69.054403}{Phys. Rev. B
  {\bf 69}, 054403}~(2004).

\bibitem{PhysRevLett.52.1}
O.~Bohigas, M.~J. Giannoni, and C.~Schmit.
\newblock ``Characterization of chaotic quantum spectra and universality of
  level fluctuation laws''.
\newblock \href{https://dx.doi.org/10.1103/PhysRevLett.52.1}{Phys. Rev. Lett.
  {\bf 52}, 1--4}~(1984).

\bibitem{PhysRevA.42.2431}
C.~H. Lewenkopf.
\newblock ``Limits of level-spacing fluctuations as a characterization of
  quantum chaos''.
\newblock \href{https://dx.doi.org/10.1103/PhysRevA.42.2431}{Phys. Rev. A {\bf
  42}, 2431--2433}~(1990).

\bibitem{Li_2008}
Hui Li and F.~D.~M. Haldane.
\newblock ``Entanglement spectrum as a generalization of entanglement entropy:
  Identification of topological order in non-abelian fractional quantum hall
  effect states''.
\newblock \href{https://dx.doi.org/10.1103/PhysRevLett.101.010504}{Phys. Rev.
  Lett. {\bf 101}, 010504}~(2008).

\bibitem{Zhang_2020}
Lei Zhang, Justin~A. Reyes, Stefanos Kourtis, Claudio Chamon, Eduardo~R.
  Mucciolo, and Andrei~E. Ruckenstein.
\newblock ``Nonuniversal entanglement level statistics in projection-driven
  quantum circuits''.
\newblock \href{https://dx.doi.org/10.1103/PhysRevB.101.235104}{Phys. Rev. B
  {\bf 101}, 235104}~(2020).

\bibitem{stabspec}
Albert~T. Schmitz, Sheng-Jie Huang, and Abhinav Prem.
\newblock ``Entanglement spectra of stabilizer codes: A window into gapped
  quantum phases of matter''.
\newblock \href{https://dx.doi.org/10.1103/PhysRevB.99.205109}{Phys. Rev. B
  {\bf 99}, 205109}~(2019).

\bibitem{True_2022}
Sarah True and Alioscia Hamma.
\newblock ``Transitions in entanglement complexity in random circuits''.
\newblock \href{https://dx.doi.org/10.22331/q-2022-09-22-818}{Quantum {\bf 6},
  818}~(2022).

\bibitem{tirrito2023quantifying}
Emanuele Tirrito, Poetri~Sonya Tarabunga, Gugliemo Lami, Titas Chanda, Lorenzo
  Leone, Salvatore F.~E. Oliviero, Marcello Dalmonte, Mario Collura, and
  Alioscia Hamma.
\newblock ``Quantifying nonstabilizerness through entanglement spectrum
  flatness''.
\newblock \href{https://dx.doi.org/10.1103/PhysRevA.109.L040401}{Phys. Rev. A
  {\bf 109}, L040401}~(2024).

\bibitem{entspecirrev}
Daniel Shaffer, Claudio Chamon, Alioscia Hamma, and Eduardo~R Mucciolo.
\newblock ``Irreversibility and entanglement spectrum statistics in quantum
  circuits''.
\newblock \href{https://dx.doi.org/10.1088/1742-5468/2014/12/p12007}{Journal of
  Statistical Mechanics: Theory and Experiment {\bf 2014}, P12007}~(2014).

\bibitem{irrev}
Claudio Chamon, Alioscia Hamma, and Eduardo~R. Mucciolo.
\newblock ``Emergent irreversibility and entanglement spectrum statistics''.
\newblock \href{https://dx.doi.org/10.1103/PhysRevLett.112.240501}{Phys. Rev.
  Lett. {\bf 112}, 240501}~(2014).

\bibitem{ShiyuChamon}
Shiyu Zhou, Zhi-Cheng Yang, Alioscia Hamma, and Claudio Chamon.
\newblock ``{Single T gate in a Clifford circuit drives transition to universal
  entanglement spectrum statistics}''.
\newblock \href{https://dx.doi.org/10.21468/SciPostPhys.9.6.087}{SciPost Phys.
  {\bf 9}, 087}~(2020).

\bibitem{TerhalDivincenzo}
Barbara~M. Terhal and David~P. DiVincenzo.
\newblock ``Classical simulation of noninteracting-fermion quantum circuits''.
\newblock \href{https://dx.doi.org/10.1103/PhysRevA.65.032325}{Phys. Rev. A
  {\bf 65}, 032325}~(2002).

\bibitem{JozsaMiyake}
Richard Jozsa and Akimasa Miyake.
\newblock ``Matchgates and classical simulation of quantum circuits''.
\newblock \href{https://dx.doi.org/10.1098/rspa.2008.0189}{Proceedings of the
  Royal Society A: Mathematical, Physical and Engineering Sciences {\bf 464},
  3089--3106}~(2008).

\bibitem{Brod_2016}
Daniel~J. Brod.
\newblock ``Efficient classical simulation of matchgate circuits with
  generalized inputs and measurements''.
\newblock \href{https://dx.doi.org/10.1103/PhysRevA.93.062332}{Phys. Rev. A
  {\bf 93}, 062332}~(2016).

\bibitem{MatchwOutside}
M.~Hebenstreit, R.~Jozsa, B.~Kraus, and S.~Strelchuk.
\newblock ``Computational power of matchgates with supplementary resources''.
\newblock \href{https://dx.doi.org/10.1103/PhysRevA.102.052604}{Phys. Rev. A
  {\bf 102}, 052604}~(2020).

\bibitem{JoszaCliffC}
Richard Jozsa and Marrten Van Den~Nest.
\newblock ``Classical simulation complexity of extended clifford circuits''.
\newblock Quantum Info. Comput. {\bf 14}, 633–648~(2014).
\newblock  url:~\url{https://doi.org/10.48550/arXiv.1305.6190}.

\bibitem{Meh2004}
Madan~Lal Mehta.
\newblock ``Random matrices''.
\newblock Academic Press. ~(2004).
\newblock 3rd edition.
\newblock  url:~\url{https://doi.org/10.1016/C2009-0-22297-5}.

\bibitem{Livan_2018}
Giacomo Livan, Marcel Novaes, and Pierpaolo Vivo.
\newblock ``Introduction to random matrices''.
\newblock \href{https://dx.doi.org/10.1007/978-3-319-70885-0}{Springer
  International Publishing}. ~(2018).

\bibitem{Atas_2013}
Y.~Y. Atas, E.~Bogomolny, O.~Giraud, and G.~Roux.
\newblock ``Distribution of the ratio of consecutive level spacings in random
  matrix ensembles''.
\newblock \href{https://dx.doi.org/10.1103/PhysRevLett.110.084101}{Phys. Rev.
  Lett. {\bf 110}, 084101}~(2013).

\bibitem{intfermions}
Vadim Oganesyan and David~A. Huse.
\newblock ``Localization of interacting fermions at high temperature''.
\newblock \href{https://dx.doi.org/10.1103/PhysRevB.75.155111}{Phys. Rev. B
  {\bf 75}, 155111}~(2007).

\bibitem{coeurjolly2006normalized}
J.~F. Coeurjolly, R.~Drouilhet, and J.~F. Robineau.
\newblock ``Normalized information-based divergences''.
\newblock \href{https://dx.doi.org/10.1134/S0032946007030015}{Problems of
  Information Transmission {\bf 43}, 167--189}~(2007).

\bibitem{nielsen00}
Michael~A. Nielsen and Isaac~L. Chuang.
\newblock ``Quantum computation and quantum information''.
\newblock Cambridge University Press. ~(2010).
\newblock 10th anniversay edition.
\newblock  url:~\url{https://doi.org/10.1017/CBO9780511976667}.

\bibitem{gottesmanknill}
D~Gottesman.
\newblock ``The heisenberg representation of quantum computers''.
\newblock \href{https://dx.doi.org/10.48550/arXiv.quant-ph/9807006}{Technical
  report}.
\newblock Los Alamos~(1998).

\bibitem{CTab}
Scott Aaronson and Daniel Gottesman.
\newblock ``Improved simulation of stabilizer circuits''.
\newblock \href{https://dx.doi.org/10.1103/PhysRevA.70.052328}{Phys. Rev. A
  {\bf 70}, 052328}~(2004).

\bibitem{Bravyi_2021}
Sergey Bravyi and Dmitri Maslov.
\newblock ``Hadamard-free circuits expose the structure of the clifford
  group''.
\newblock \href{https://dx.doi.org/10.1109/tit.2021.3081415}{{IEEE}
  Transactions on Information Theory {\bf 67}, 4546--4563}~(2021).

\bibitem{Valiant}
Leslie~G. Valiant.
\newblock ``Quantum computers that can be simulated classically in polynomial
  time''.
\newblock In Proceedings of the Thirty-Third Annual ACM Symposium on Theory of
  Computing.
\newblock \href{https://dx.doi.org/10.1145/380752.380785}{Page 114–123}.
\newblock STOC '01New York, NY, USA~(2001). Association for Computing
  Machinery.

\bibitem{knill2001fermionic}
E.~Knill.
\newblock ``Fermionic linear optics and matchgates''~(2001).
\newblock
  \href{http://arxiv.org/abs/quant-ph/0108033}{arXiv:quant-ph/0108033}.

\bibitem{Brod_2012}
Daniel~J. Brod and Ernesto~F. Galv\~ao.
\newblock ``Geometries for universal quantum computation with matchgates''.
\newblock \href{https://dx.doi.org/10.1103/PhysRevA.86.052307}{Phys. Rev. A
  {\bf 86}, 052307}~(2012).

\bibitem{BrodXY}
Daniel~J. Brod and Andrew~M. Childs.
\newblock ``The computational power of matchgates and the xy interaction on
  arbitrary graphs''.
\newblock Quantum Info. Comput. {\bf 14}, 901–916~(2014).
\newblock  url:~\url{https://doi.org/10.26421/QIC14.11-12-1}.

\bibitem{Bravyi_2002}
Sergey~B. Bravyi and Alexei~Yu. Kitaev.
\newblock ``Fermionic quantum computation''.
\newblock \href{https://dx.doi.org/10.1006/aphy.2002.6254}{Annals of Physics
  {\bf 298}, 210–226}~(2002).

\bibitem{PhysRev.80.268}
G.~C. Wick.
\newblock ``The evaluation of the collision matrix''.
\newblock \href{https://dx.doi.org/10.1103/PhysRev.80.268}{Phys. Rev. {\bf 80},
  268--272}~(1950).

\bibitem{Surace_2022}
Jacopo Surace and Luca Tagliacozzo.
\newblock ``Fermionic gaussian states: an introduction to numerical
  approaches''.
\newblock \href{https://dx.doi.org/10.21468/scipostphyslectnotes.54}{SciPost
  Physics Lecture Notes}~(2022).

\bibitem{chien2020custom}
Riley~W. Chien and James~D. Whitfield.
\newblock ``Custom fermionic codes for quantum simulation''~(2020).
\newblock  \href{http://arxiv.org/abs/2009.11860}{arXiv:2009.11860}.

\bibitem{1928ZPhy...47..631J}
P.~{Jordan} and E.~{Wigner}.
\newblock ``{{\"U}ber das Paulische {\"A}quivalenzverbot}''.
\newblock \href{https://dx.doi.org/10.1007/BF01331938}{Zeitschrift fur Physik
  {\bf 47}, 631--651}~(1928).

\bibitem{Jozsa_2009}
Richard Jozsa, Barbara Kraus, Akimasa Miyake, and John Watrous.
\newblock ``Matchgate and space-bounded quantum computations are equivalent''.
\newblock \href{https://dx.doi.org/10.1098/rspa.2009.0433}{Proceedings of the
  Royal Society A: Mathematical, Physical and Engineering Sciences {\bf 466},
  809–830}~(2009).

\bibitem{Ji2019Nov}
Jia-Wei Ji and David~L. Feder.
\newblock ``{Extending matchgates to universal quantum computation via the
  Hubbard model}''.
\newblock \href{https://dx.doi.org/10.1103/PhysRevA.100.052324}{Phys. Rev. A
  {\bf 100}, 052324}~(2019).

\bibitem{watrous2008quantum}
John Watrous.
\newblock ``Quantum computational complexity''~(2008).
\newblock  \href{http://arxiv.org/abs/0804.3401}{arXiv:0804.3401}.

\bibitem{Whitfield_2013}
James~Daniel Whitfield, Peter~John Love, and Alán Aspuru-Guzik.
\newblock ``Computational complexity in electronic structure''.
\newblock \href{https://dx.doi.org/10.1039/c2cp42695a}{Phys. Chem. Chem. Phys.
  {\bf 15}, 397–411}~(2013).

\bibitem{aaronson2009bqp}
Scott Aaronson.
\newblock ``{BQP} and the polynomial hierarchy''~(2009).
\newblock  \href{http://arxiv.org/abs/0910.4698}{arXiv:0910.4698}.

\bibitem{Bennett_1997}
Charles~H. Bennett, Ethan Bernstein, Gilles Brassard, and Umesh Vazirani.
\newblock ``Strengths and weaknesses of quantum computing''.
\newblock \href{https://dx.doi.org/10.1137/s0097539796300933}{SIAM Journal on
  Computing {\bf 26}, 1510–1523}~(1997).

\bibitem{Brod_2011}
Daniel~J. Brod and Ernesto~F. Galvão.
\newblock ``Extending matchgates into universal quantum computation''.
\newblock \href{https://dx.doi.org/10.1103/physreva.84.022310}{Physical Review
  A{\bf 84}}~(2011).

\bibitem{mocherla2024extending}
Avinash Mocherla, Lingling Lao, and Dan~E. Browne.
\newblock ``Extending matchgate simulation methods to universal quantum
  circuits''~(2024).
\newblock  \href{http://arxiv.org/abs/2302.02654}{arXiv:2302.02654}.

\bibitem{PhysRevLett.71.1291}
Don~N. Page.
\newblock ``Average entropy of a subsystem''.
\newblock \href{https://dx.doi.org/10.1103/PhysRevLett.71.1291}{Phys. Rev.
  Lett. {\bf 71}, 1291--1294}~(1993).

\bibitem{Liu_2018}
Zi-Wen Liu, Seth Lloyd, Elton Zhu, and Huangjun Zhu.
\newblock ``Entanglement, quantum randomness, and complexity beyond
  scrambling''.
\newblock \href{https://dx.doi.org/10.1007/JHEP07(2018)041}{Journal of High
  Energy Physics {\bf 2018}, 41}~(2018).

\bibitem{jozsa2015jordanwigner}
Richard Jozsa, Akimasa Miyake, and Sergii Strelchuk.
\newblock ``Jordan-wigner formalism for arbitrary 2-input 2-output matchgates
  and their classical simulation''~(2015).
\newblock  \href{http://arxiv.org/abs/1311.3046}{arXiv:1311.3046}.

\bibitem{henderson2024hartreefockbogoliubov}
Thomas~M. Henderson, Shadan~Ghassemi Tabrizi, Guo~P. Chen, and Gustavo~E.
  Scuseria.
\newblock ``Hartree-fock-bogoliubov theory for number-parity--violating
  fermionic hamiltonians''~(2024).
\newblock  \href{http://arxiv.org/abs/2311.11553}{arXiv:2311.11553}.

\bibitem{Chapman_2020}
Adrian Chapman and Steven~T. Flammia.
\newblock ``Characterization of solvable spin models via graph invariants''.
\newblock \href{https://dx.doi.org/10.22331/q-2020-06-04-278}{Quantum {\bf 4},
  278}~(2020).

\bibitem{chapman2023unified}
Adrian Chapman, Samuel~J. Elman, and Ryan~L. Mann.
\newblock ``A unified graph-theoretic framework for free-fermion
  solvability''~(2023).
\newblock  \href{http://arxiv.org/abs/2305.15625}{arXiv:2305.15625}.

\bibitem{PhysRevB.100.214301}
Jason Iaconis, Sagar Vijay, and Rahul Nandkishore.
\newblock ``Anomalous subdiffusion from subsystem symmetries''.
\newblock \href{https://dx.doi.org/10.1103/PhysRevB.100.214301}{Phys. Rev. B
  {\bf 100}, 214301}~(2019).

\bibitem{Iaconis_2021}
Jason Iaconis.
\newblock ``Quantum state complexity in computationally tractable quantum
  circuits''.
\newblock \href{https://dx.doi.org/10.1103/PRXQuantum.2.010329}{PRX Quantum
  {\bf 2}, 010329}~(2021).

\end{thebibliography}

\onecolumn

\newpage
\newpage
\appendix

\section*{\huge Appendices}

\section{Comparisons between Matchgate plus SWAP and Clifford plus $T$}{\label{app:ClifC}}

\indent In Zhou et al.\ \cite{ShiyuChamon}, when studying Clifford plus $T$ circuits the authors are able to fit the convergence of $\delta r$ to $N\cdot N_{T}$, as their curves collapse to a single universal scaling function. While we can attempt to do the same for matchgate circuits, the fitting is less universal due to a finite size effect. \\

\begin{figure}[h]
    \begin{subfigure}{0.65\textwidth}
    	\hspace{3mm}
        \includegraphics[width=.65\linewidth]{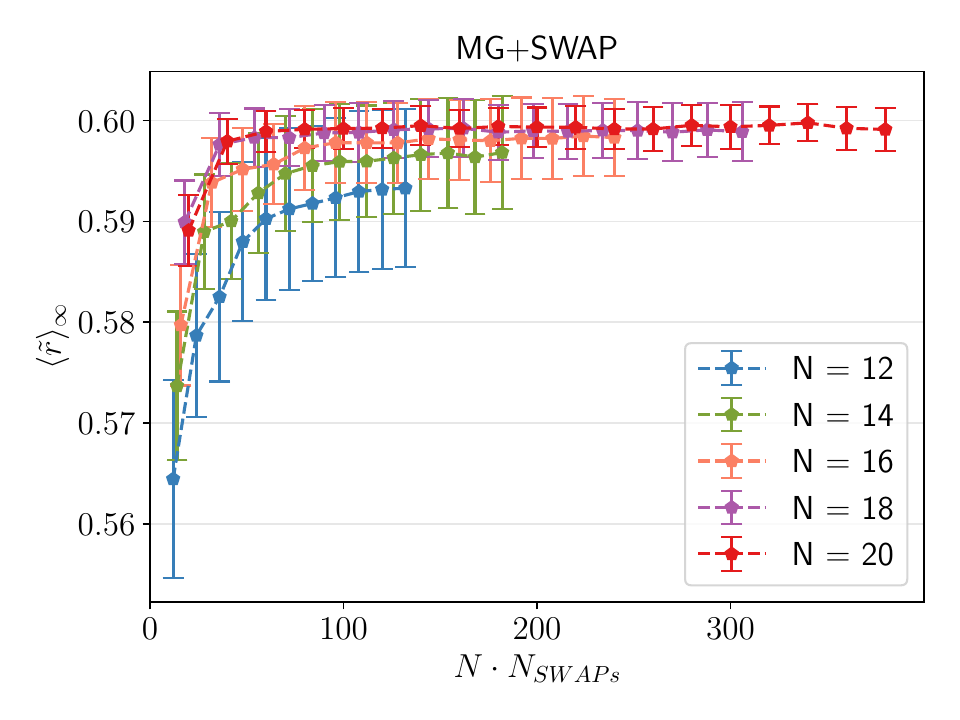}
    \end{subfigure}%
    ~ 
    \begin{subfigure}{0.65\textwidth}
        \hspace{-2cm}
        \includegraphics[width=.65\linewidth]{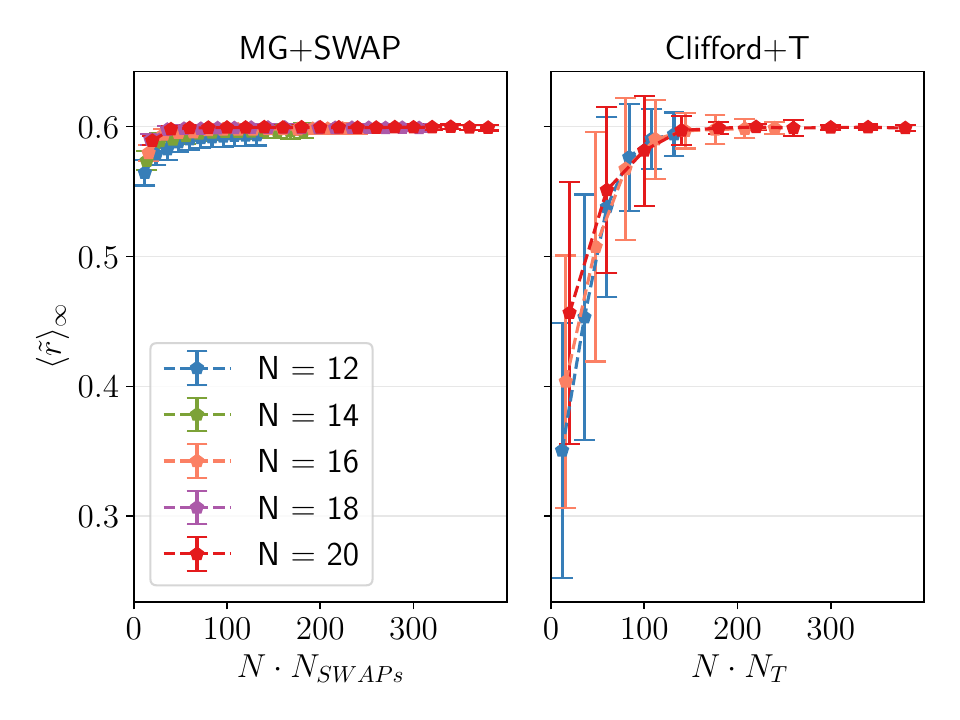}
    \end{subfigure}
	\caption{Plotting of $\langle \tilde r \rangle_{\infty}$ as a function of $N\cdot N_{\textrm SWAP}$, independently as well as scaled against $\langle \tilde r \rangle_{\infty}$ for the case of Clifford plus $T$ gates. While for the Clifford plus $T$ case a single smooth function is interpolated to, for matchgate plus SWAP depending on system size there is a finite size effect and values plateau below the true GUE value. } 
	\label{fig:CompFig}
\end{figure}

As seen in Fig.\ \ref{fig:CompFig}, past a certain number of SWAPs injected, there is no change in $\langle \tilde r \rangle$, and the value of $\langle \tilde r \rangle$ converges below the value for the Wigner-Dyson distribution. When scaled against the Clifford plus $T$ case, the scale removes some variation even though there is still a visible difference. The difference can be understood by observing that, for small systems and a small number of $T$ gates, $\langle \tilde r \rangle_{\infty}$ is well below the Wigner-Dyson value. However, even for small systems, a single SWAP gate injected in a matchgate circuit is nearly enough to reach the $\langle \tilde r \rangle \approx 0.6$ value. The scaling in the matchgate with SWAP circuits reveals a fundamental result: for small system sizes, even circuits that have a large amount of universal resources have an entanglement spectrum that is not quite Wigner-Dyson. \\

\begin{figure}[htpb]
\centering
    \hspace{-6mm}
    \includegraphics[width=0.6\linewidth]{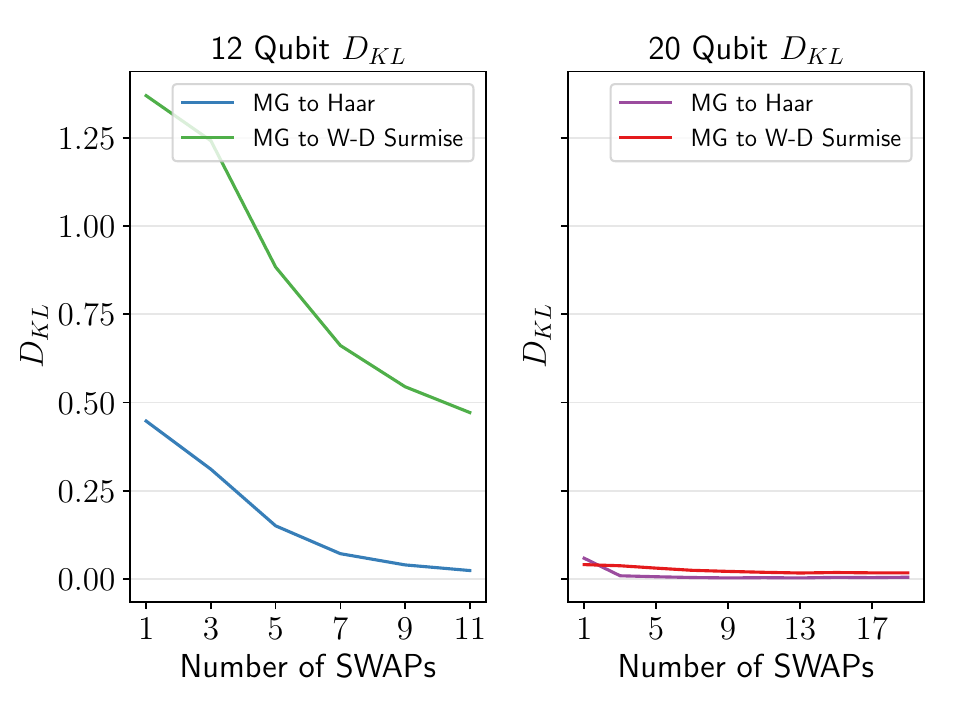}
    \caption{Kullback-Leibler divergence for 12 and 20 qubits matchgate circuits with added SWAPs. For the 12-qubit system, while adding SWAPs decreases the divergence from the Wigner-Dyson distribution on 12 qubits, there is still a significant distance to the Wigner-Dyson distribution. For 20 qubits, this distance to the Wigner-Dyson distribution closes.}
    \label{fig:KLD}
\end{figure}
\indent The explanation for this finite size effect has to do with the convergence to GUE entanglement spectrum statistics as seen in Haar random states of various sizes. Using the Kullback-Leibler divergence, we can examine the difference in the entanglement spectrum distribution with matchgates plus SWAPs between the Wigner-Dyson distribution and the entanglement spectrum distribution for Haar brickwork circuits of different system sizes. As seen in Fig.\ \ref{fig:KLD} for 12 qubits, while adding SWAPs slowly causes the matchgate distributions to converge to the 12-qubit Haar brickwork distributions, there is still a significant difference to the Wigner-Dyson distribution. For 20 qubits, even with minimal SWAPs added the difference between the matchgate distributions and Wigner-Dyson distribution is near zero. For small systems, the addition of a universal gate element does not make the distribution converge to the Wigner-Dyson distribution, but instead to the distribution of a Haar random state, which for small system sizes is itself different from the true Wigner-Dyson distribution.\\

\begin{figure}[t!]
    \begin{subfigure}{0.65\textwidth}
    	\hspace{3mm}
        \includegraphics[width=.65\linewidth]{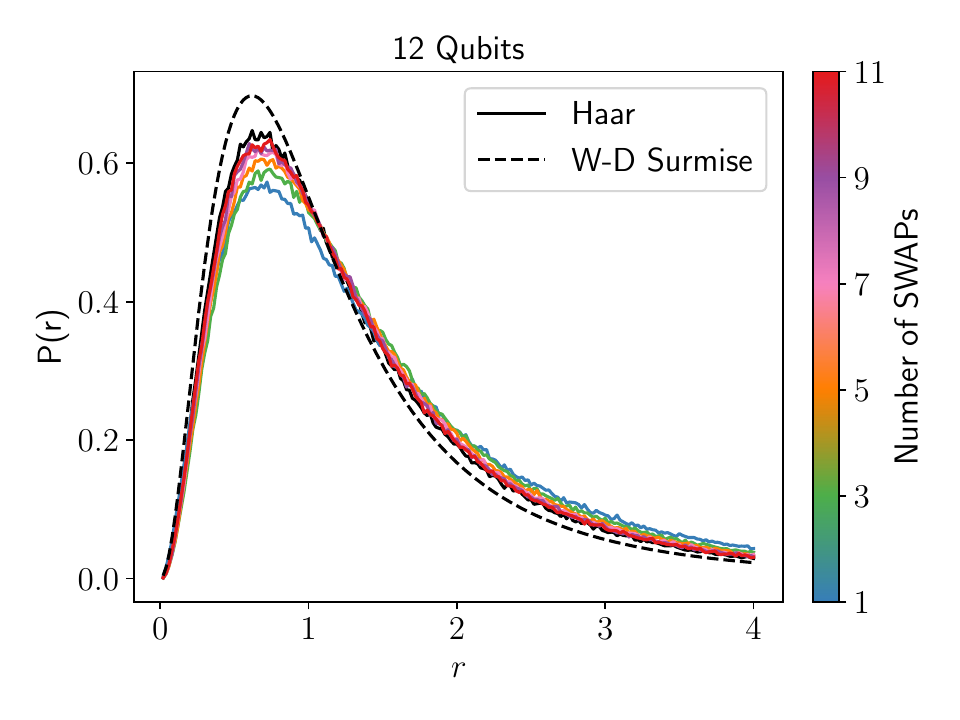}
    \end{subfigure}%
    ~ 
    \begin{subfigure}{0.65\textwidth}
        \hspace{-2cm}
        \includegraphics[width=.65\linewidth]{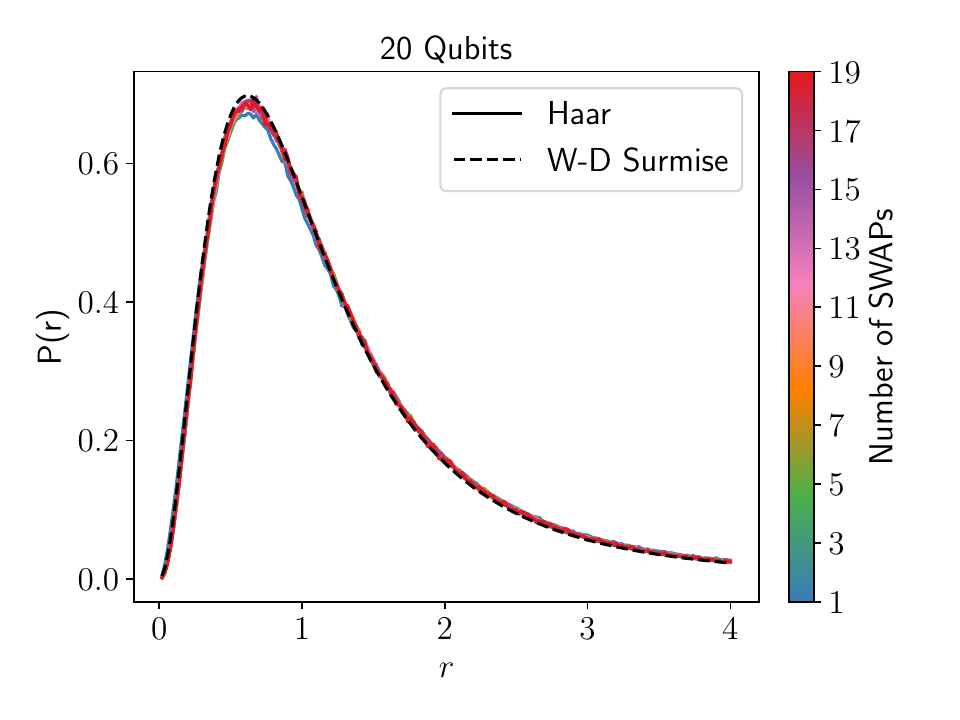}
    \end{subfigure}
	\caption{Distribution of level spacing for matchgate circuits of 12 and 20 qubits with injected SWAPs, compared to both a Haar brickwork circuit of 12 and 20 qubits, and the Wigner-Dyson distribution. For 12 qubits we can see that while for many SWAPs the matchgate circuit distributions converge near the distribution for a 12 qubit Haar state, there is still distance between the Haar brickwork distribution and the Wigner-Dyson distribution. This distance closes though as we observe the entanglement spectrum distributions for system sizes of $20$. } 
	\label{fig:PrFig}
\end{figure}

\indent We can also see the disjoint between the entanglement spectrum distribution in small Haar random states versus the GUE distribution by explicitly plotting the distributions $P(r)$ as a function of $r$ for our systems. As seen in Fig.\ \ref{fig:PrFig}, while for 12 qubits the addition of even 11 SWAPs converges to the entanglement spectrum distribution for Haar distributed states, even the 12-qubit Haar distributed state has a distribution that doesn't overlap with the Wigner-Dyson distribution. As we increase system size to 20, the distribution from the Haar random state is in near perfect agreement with the GUE distribution, and we find that matchgate circuits with a small number of injected SWAPs are near the Wigner-Dyson distribution. 

\section{Entropy with Injected SWAPs and Varied Input States}\label{App:Ent}

\begin{figure}[htpb]
    \centering
    \includegraphics[width=0.6\linewidth]{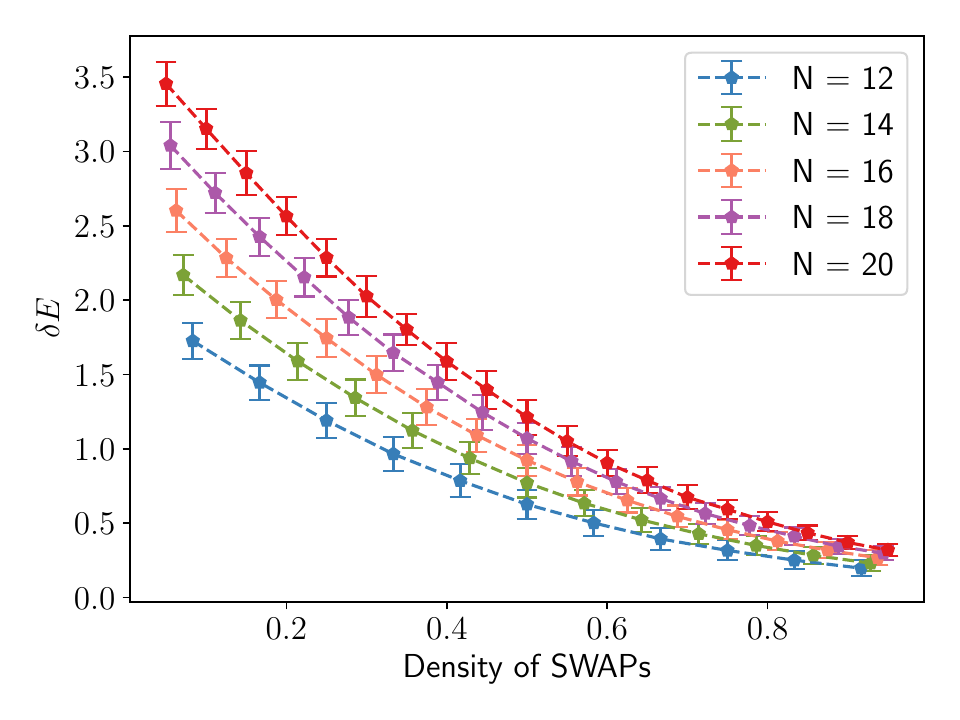}
    \caption{The difference of average entropy in matchgate circuits with density of injected SWAPs to the Page entropy respective of system size. We find that though all curves start at a different point, they behave in similar ways, in which as the density of SWAPs increases towards one the entropy in the matchgate circuit approaches the Page entropy. }
    \label{fig:totent}
\end{figure}

\indent We study the entropy of matchgate circuits post-SWAP injection in the infinite time regime, in Fig.\ \ref{fig:totent}. We find that as a function of density of SWAPs in the system, matchgate circuits with injected SWAPs approach the Page entropy. \\

\begin{figure}
    \centering
    \includegraphics[width=0.6\linewidth]{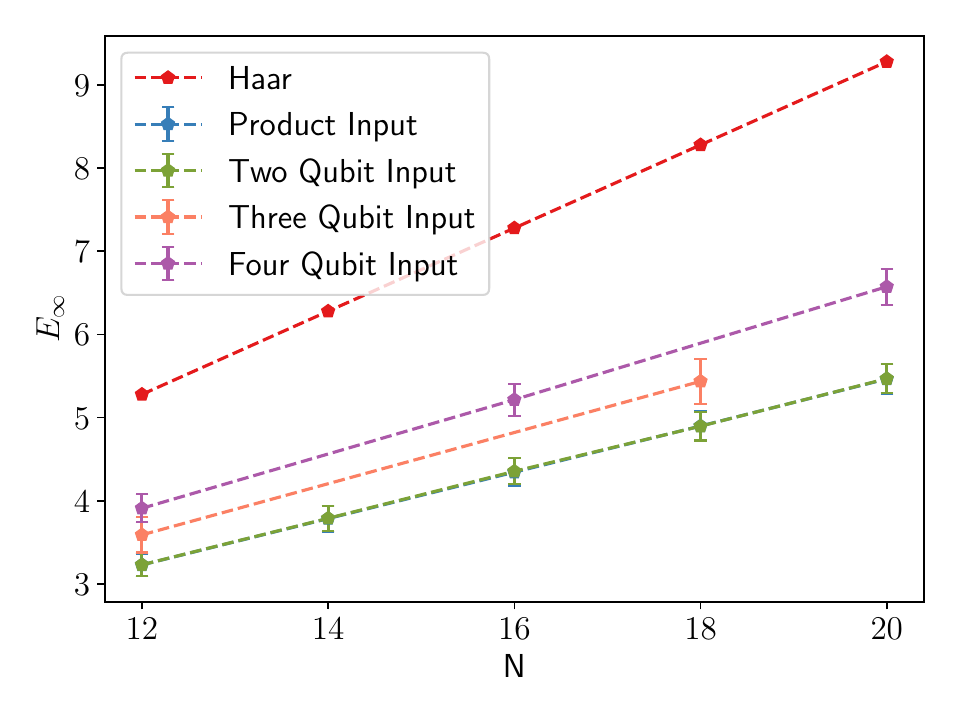}
    \caption{Entanglement as a function of system size for matchgate circuits on different sized inputs. The entanglement for product and two-qubit inputs is near identical. } 
    \label{fig:entinputs}
\end{figure}

\indent Tracking the entanglement of matchgate circuits with varying inputs, as seen in Fig.\ \ref{fig:entinputs}, shows us that as in the case with injected SWAPs, while having three or four qubit entangled input states leads to complex entanglement, the sharp jump in the complexity of entanglement does not manifest in the single value of entanglement, as circuits with neither three or four qubit entangled state inputs approaches the Page entropy as a function of $N$. It also shows us that not only does the entanglement complexity for product and two-qubit inputs overlap, but the entanglement does too. \\

\begin{figure}[htpb]
    \centering
    \includegraphics[width=0.6\linewidth]{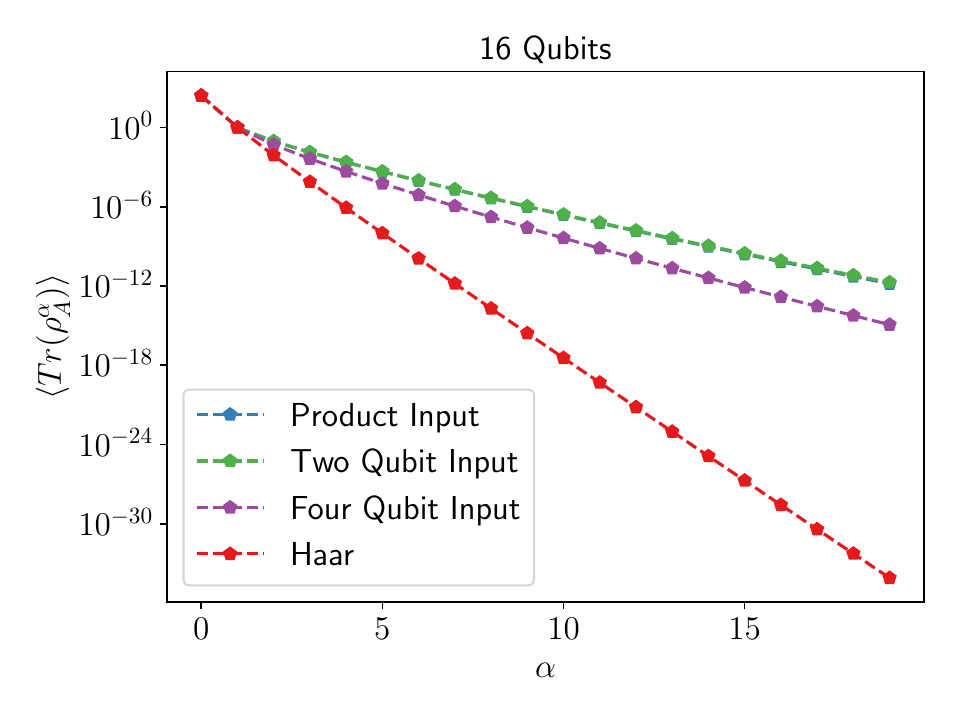}
    \caption{ $\langle Tr(\rho_A^\alpha)\rangle$ for matchgates circuits of $N=16$, and varying input states, compared to the same for evolution under Haar brickwork circuits.} 
    \label{fig:renyiiinput}
\end{figure}

\indent Further than just entanglement, we can examine more generalized entropies by studying the R\'enyi entropies $R_\alpha = \frac{1}{1-\alpha}\log{\sum_i p_i^{\alpha}}$, and the trace of the reduced density matrix to power $k$. In Liu et al.\ \cite{Liu_2018}, the authors discuss the connection between unitary $t$-designs and the R\'enyi entropy, and find that the $\alpha$-R\'enyi averaged over a unitary $\alpha$-design is near maximal. Thus, R\'enyi entropies can give insight into the complexity of the wave function, compared to Haar random wave functions. They also show that, for some $\alpha$-design, that 
\begin{equation}
    \langle Tr(\rho_A^\alpha)\rangle_{\alpha} = \langle Tr(\rho_A^\alpha) \rangle_{Haar}
\end{equation}
Said otherwise, for some unitary $\alpha$-design, the trace of $\rho_A^{\alpha}$ is equal to that of Haar random unitaries. \\
\indent As seen in Fig.\ \ref{fig:renyiiinput}, $\langle Tr(\rho_A^\alpha)\rangle$ is identical for product and two-qubit entangled inputs, though for four-qubit entangled inputs $\langle Tr(\rho_A^\alpha)\rangle$ is closer to matchgates than it is to the metric for the Haar distribution. Quantifying this difference between the two more rigorously is the topic of future work. \\

\section{Conjecture on Clifford-conjugated Matchgate Circuits}\label{App:Conj}

\indent While the analysis on why some simulable Clifford and matchgate hybrid circuits have Wigner-Dyson distributed entanglement spectra while other do not is an open problem, we conjecture that in the thermodynamic limit, every such Clifford that conjugates our fermionic operators such that $Z_k$ can be written as a product of a maximum of two operators leads to Poisson entanglement spectra, while every Clifford circuit such that conjugation leads to $Z_k$ being a product of a maximum of greater than $2$ fermionic operators (but not $O(n)$, so that circuits remain simulable) leads to a Wigner-Dyson entanglement spectrum. \\
\indent As an example, we will take take two Clifford circuits 
\begin{subequations}
\begin{align}
    C_3 &= H_1 H_2 CNOT_{12}\\
    C_4 &=  H_1 H_2 H_3 CNOT_{12} CNOT_{23}
\end{align}
\end{subequations}
In both circuits, we only conjugate two or three qubits non-trivially: for $C_3$ we retain $Z_k$ being written as two fermionic operators, while in $C_4$ we have that a single $Z_k$ term is written as the product of four fermionic operators. \\

\begin{figure}
    \centering
    \includegraphics[width=0.6\linewidth]{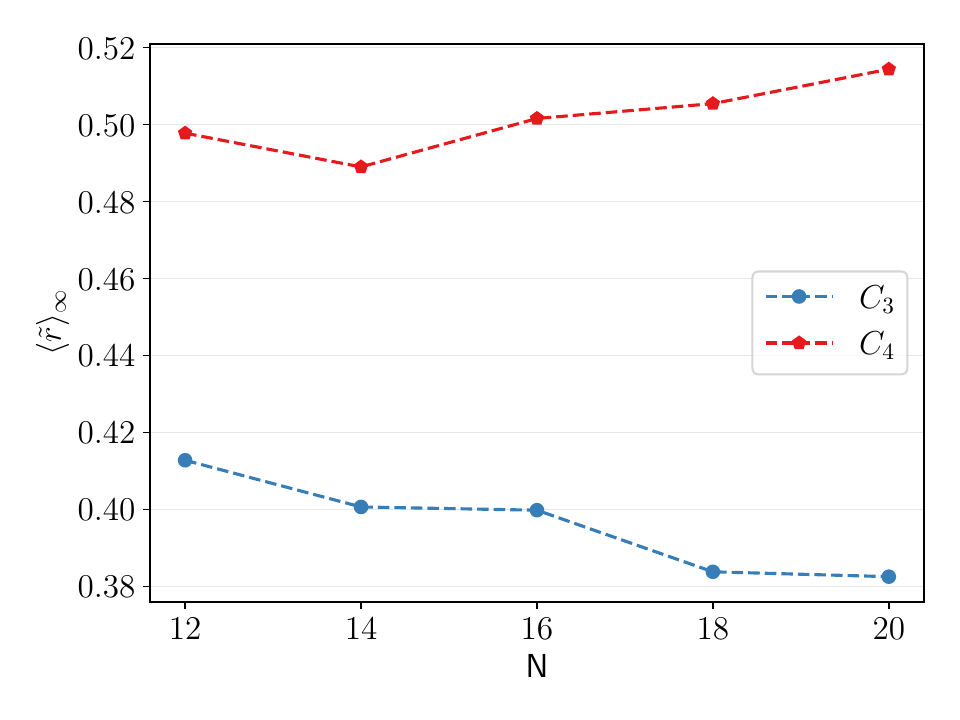}
    \caption{Comparisons in $\langle \tilde r \rangle$ for circuits on product state inputs as a function of system size, for matchgate circuits conjugated by $C_3$ and $C_4$. While for $C_3$, in which $Z_k$ remains a product of only two fermionic operators, for $C_4$ it seems that as a function of system size, $\langle \tilde r \rangle$ deviates further from the Poisson value and closer to Wigner-Dyson. More computational power is needed to numerically simulate greater system sizes, and see the large $N$ behavior for $C_4$. } 
    \label{fig:t3t4}
\end{figure}

\indent As seen in Fig.\ \ref{fig:t3t4}, while conjugation by $C_3$ produces the same behavior as how matchgate circuits behaved prior to SWAP insertion, conjugation by $C_4$ produces $\langle \tilde r \rangle$ that seems to slowly grow as a function of system size. There thus seems to be a connection between the entanglement spectrum and how many terms $Z_k$ can be expressed in. Future work involves proving this conjecture, with work involving understanding every fermion-to-qubit encoding such that $Z_k$ is written as more than two fermionic operators, but not $O(n)$ of them. \\


\end{document}